\documentclass[sigconf,nonacm]{acmart}
\newcommand{\pagefrac}[1]{\dimexpr\ifdim#1\textwidth>\linewidth\linewidth\else#1\textwidth\fi\relax}

\AtBeginDocument{%
  }

\setcopyright{acmlicensed}
\copyrightyear{2026}
\acmYear{2026}
\acmDOI{XXXXXXX.XXXXXXX}
\acmConference[Preprint]{Preprint}{2026}{arXiv}
\acmISBN{978-1-4503-XXXX-X/2026/06}

\usepackage{float}
\usepackage[htt]{hyphenat}
\usepackage{tabularx}

\usepackage{tikz}
\usetikzlibrary{arrows.meta}
\usepackage{xspace}
\definecolor{ConceptColor}{HTML}{334155}   %
\definecolor{IncreaseColor}{HTML}{15803D}  %
\definecolor{DecreaseColor}{HTML}{B91C1C}  %
\definecolor{RelationColor}{HTML}{57534E}  %
\newcommand{\chip}[2]{%
  \tikz[baseline=(chip.base)]{\node[inner xsep=3pt, inner ysep=1.2pt,
    rounded corners=2.5pt, fill=#1!8, draw=#1!30, text=#1!85!black,
    font=\footnotesize\ttfamily] (chip) {#2};}\xspace}
\newcommand{\cpt}[1]{\chip{ConceptColor}{#1}}

\newcommand{\edgelabel}[2]{%
  \hspace{0.15em}\tikz[baseline=(r.base)]{%
    \node[inner sep=0pt, outer sep=0pt, font=\scriptsize\ttfamily,
          text=#1!85!black] (r) {#2};
    \draw[#1!55, line width=0.45pt] ([xshift=-4pt]r.west) -- (r.west);
    \draw[#1!55, line width=0.45pt, -{Stealth[length=3.2pt,width=2.6pt]}]
         (r.east) -- ([xshift=5pt]r.east);%
  }\hspace{0.15em}}
\newcommand{\edgeup}[1]{\edgelabel{IncreaseColor}{#1}}
\newcommand{\edgedn}[1]{\edgelabel{DecreaseColor}{#1}}

\newcommand{\triple}[3]{\cpt{#1}#2\cpt{#3}}

\usepackage{listings}
\lstdefinestyle{prompt}{
  basicstyle=\scriptsize\ttfamily,
  breaklines=true,
  breakautoindent=false,
  breakindent=0pt,
  columns=fullflexible,
  keepspaces=true,
  showstringspaces=false,
  frame=none,
  xleftmargin=0pt,
  aboveskip=0.5\baselineskip,
  belowskip=0.5\baselineskip,
  extendedchars=true,
  inputencoding=utf8,
  literate=%
    {—}{{---}}3 {–}{{--}}2 {…}{{...}}3
    {→}{{$\rightarrow$}}1 {←}{{$\leftarrow$}}1
    {─}{{-}}1 {│}{{|}}1 {├}{{|}}1 {└}{{`}}1
    {≤}{{$\leq$}}1 {≥}{{$\geq$}}1 {·}{{$\cdot$}}1
    {α}{{$\alpha$}}1 {²}{{$^2$}}1 {⁺}{{$^+$}}1
    {✓}{{$\checkmark$}}1 {✗}{{$\times$}}1
    {‘}{{`}}1 {’}{{'}}1 {“}{{``}}1 {”}{{''}}1
}

\newcommand{\sysname}{Asterism}
\newcommand{\persona}{Dora}

\usepackage{xcolor}

\renewcommand\fbox{\fcolorbox{gray}{white}}

\begin{document}

\setlength{\fboxrule}{0.05pt}
\setlength{\fboxsep}{0pt}

\makeatletter
\def\titlenote#1{%
  \g@addto@macro\@title{\footnotemark}%
  \g@addto@macro\@titlenotes{\stepcounter{footnote}\footnotetext{#1}}}
\makeatother
\definecolor{wildstrawberry}{rgb}{1.0, 0.26, 0.64}

\title[\sysname: Synthesizing Scattered Observations into Literature-Grounded Hypotheses and Theories]{\sysname{}: Exploring and Synthesizing Scattered Observations\\into Literature-Grounded Hypotheses and Theories}
\titlenote{An asterism is a pattern of stars that observers pick out and name for themselves, but without the official standing the astronomical community confers on its 88 constellations. The theories a researcher builds with \sysname{} are asterisms: the stars are the literature's scattered observations, the researcher draws the lines between them, and the patterns are provisional, pointing to new ideas and experiments.}

\author{Joseph Chee Chang}
\affiliation{%
  \institution{Allen Institute for AI}
  \city{Seattle}
  \state{Washington}
  \country{United States}
}
\email{josephc@allenai.org}

\author{Michael D'Arcy}
\affiliation{%
  \institution{Allen Institute for AI}
  \city{Seattle}
  \state{Washington}
  \country{United States}
}
\email{miked@allenai.org}

\author{Amy X. Zhang}
\affiliation{%
  \institution{University of Washington}
  \city{Seattle}
  \state{Washington}
  \country{United States}
}
\email{axz@cs.uw.edu}

\author{Pao Siangliulue}
\affiliation{%
  \institution{Allen Institute for AI}
  \city{Seattle}
  \state{Washington}
  \country{United States}
}
\email{paos@allenai.org}

\author{Sangho Suh}
\affiliation{%
  \institution{Allen Institute for AI}
  \city{Seattle}
  \state{Washington}
  \country{United States}
}
\email{sanghos@allenai.org}

\author{Aakanksha Naik}
\affiliation{%
  \institution{Allen Institute for AI}
  \city{Seattle}
  \state{Washington}
  \country{United States}
}
\email{aakankshan@allenai.org}

\author{Jena D. Hwang}
\affiliation{%
  \institution{Allen Institute for AI}
  \city{Seattle}
  \state{Washington}
  \country{United States}
}
\email{jenah@allenai.org}

\author{Javier Ramos Benitez}
\affiliation{%
  \institution{University of Washington}
  \city{Seattle}
  \state{Washington}
  \country{United States}
}
\email{javier98@uw.edu}

\author{Stella Wroblewski}
\affiliation{%
  \institution{Allen Institute for AI}
  \city{Seattle}
  \state{Washington}
  \country{United States}
}
\email{stellaw@allenai.org}

\author{Matt Latzke}
\affiliation{%
  \institution{Allen Institute for AI}
  \city{Seattle}
  \state{Washington}
  \country{United States}
}
\email{mattl@allenai.org}

\author{Michael Cuoco}
\affiliation{%
  \institution{Allen Institute}
  \city{Seattle}
  \state{Washington}
  \country{United States}
}
\email{mike.cuoco@alleninstitute.org}

\author{Ruben Lozano-Aguilera}
\affiliation{%
  \institution{Allen Institute for AI}
  \city{Seattle}
  \state{Washington}
  \country{United States}
}
\email{rubenl@allenai.org}

\author{Kris Ganjam}
\affiliation{%
  \institution{Allen Institute}
  \city{Seattle}
  \state{Washington}
  \country{United States}
}
\email{kris.ganjam@alleninstitute.org}

\author{Joel Chan}
\affiliation{%
  \institution{University of Maryland}
  \city{College Park}
  \state{Maryland}
  \country{United States}
}
\email{joelchan@umd.edu}

\author{Doug Downey}
\affiliation{%
  \institution{Allen Institute for AI}
  \city{Seattle}
  \state{Washington}
  \country{United States}
}
\email{dougd@allenai.org}

\author{Peter Jansen}
\affiliation{%
  \institution{Allen Institute for AI}
  \city{Seattle}
  \state{Washington}
  \country{United States}
}
\email{peterj@allenai.org}

\author{Kyle J. Travaglini}
\affiliation{%
  \institution{Allen Institute}
  \city{Seattle}
  \state{Washington}
  \country{United States}
}
\email{kyle.travaglini@alleninstitute.org}

\author{Daniel S. Weld}
\affiliation{%
  \institution{Allen Institute for AI}
  \city{Seattle}
  \state{Washington}
  \country{United States}
}
\email{danw@allenai.org}

\renewcommand{\shortauthors}{Chang et al.}

\begin{abstract}

A theory draws many independent observations into one framework with novel hypotheses. A researcher building such a theory must synthesize observations scattered across many papers, each describing related concepts but often in different terms.  Which concepts matter most also depends on their preferences and research questions. Recent approaches scale theory synthesis with LLMs, but automate away choices and intuitions from researchers. We present \sysname{}, which extracts observations from hundreds of papers as concept-relation triples, with concepts unified in a hierarchical ontology. Researchers curate an evidence graph using the ontology and aggregate observations at different levels of granularity to focus theory formation on specific phenomena of interest.  In a field deployment (n=10), researchers worked from observations to theories, and kept concepts and hypotheses fitting their preferences. In two case studies, teams of immunology and agriculture researchers discovered mechanisms outside their standard analyses and constructed hypotheses worth follow-up experiments.

\end{abstract}



\maketitle

\begin{figure*}[th]
    \centering
    \fbox{\includegraphics[width=1\linewidth]{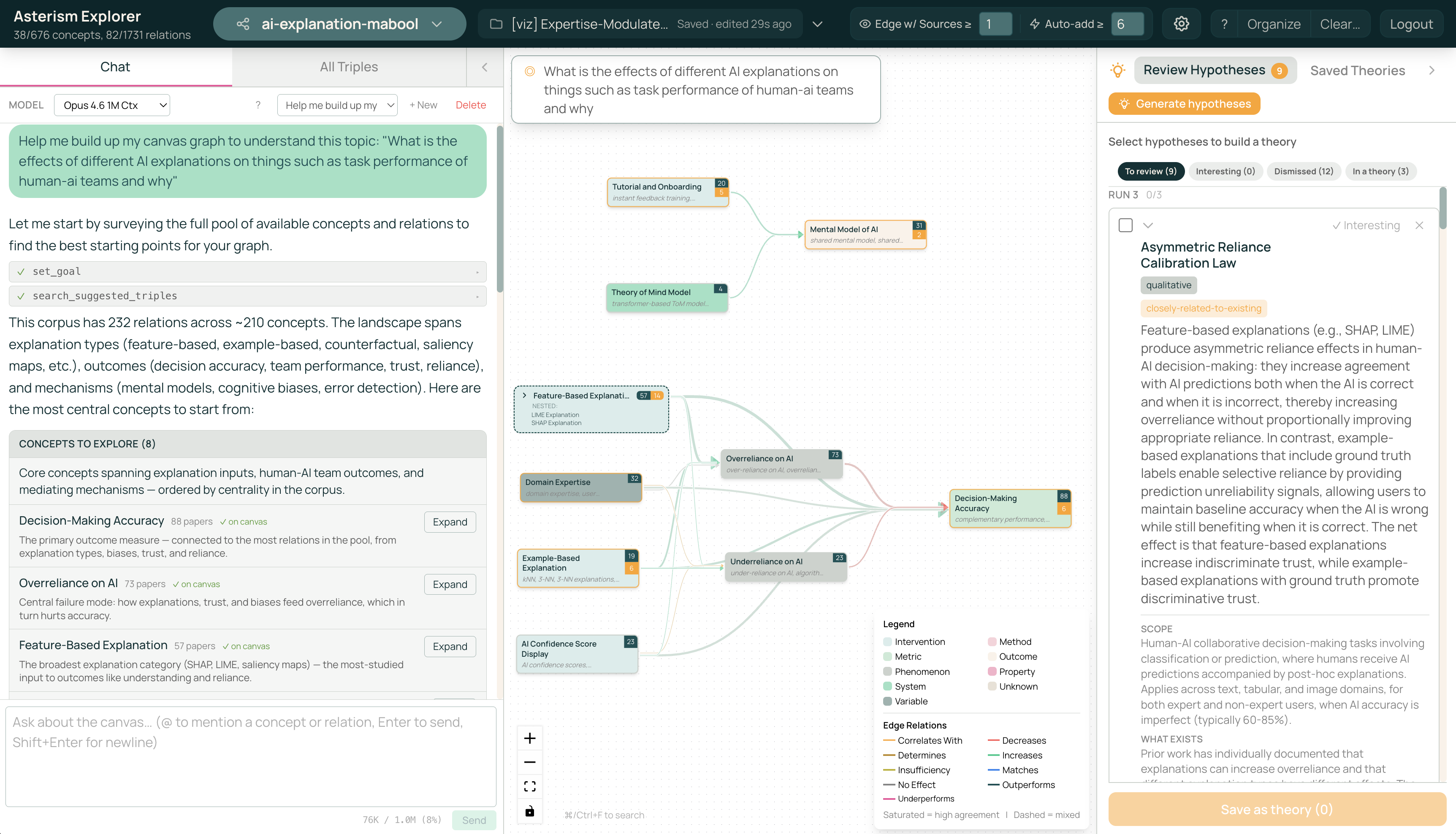}}
    \caption{\sysname{} on a graph extracted  from around 100 papers about how AI explanations affect human-AI teams. \textbf{Left:} a chat agent that searches the extracted evidence on the user's behalf, answers questions about it, and proposes extracted concepts and relations to be added to the canvas (Figure~\ref{fig:triples}). \textbf{Middle:} the canvas, holding the concepts and relations the user has curated for their inquiry, tinted by concept category and colored by relation, with a badge on every node and edge counting the unique papers behind it (Figures~\ref{fig:nesting} and~\ref{fig:node-panel}). \textbf{Right:} the hypothesis drawer, where hypotheses generated based on evidence from the canvas are reviewed by the user, marked interesting or dismissed, and selected for synthesis into a theory (Figure~\ref{fig:theory-footprint}). Each hypothesis lists the concept-relation triples it was drawn from, hovering highlights them on the canvas (Figure~\ref{fig:theory-footprint}).}
    \Description{Screenshot of the \sysname{} web interface, divided into three vertical panels. The left panel is a chat conversation in which the user asks the system to help build up a graph for a research question and the agent replies with a ranked list of concepts to explore, each with a paper count and an Add or Expand button. The middle panel is a node-link canvas holding about a dozen labeled concept boxes, such as Feature-Based Explanation, Example-Based Explanation, Domain Expertise, Overreliance on AI, Underreliance on AI, and Decision-Making Accuracy, joined by curved colored edges. The right panel lists generated hypotheses; the open one is titled Asymmetric Reliance Calibration Law and shows its statement, its scope, and its supporting evidence with concept-relation triples.}
    \label{fig:system}
\end{figure*}

\section{Introduction}

One of the fundamental goals of science is to better understand the world around us. Researchers conduct \emph{experiments} and report \emph{observations} which support \emph{claims} often about \emph{relationships between concepts}, such as how one variable increases or decreases another \cite{265689492}.\footnote{We deliberately avoid the word ``causal'' in this paper since what qualifies differs in different fields (e.g., particular experiment protocols). Our system extracts \emph{relationships} from papers in accordance with the norms of the community where the papers were published.}
Synthesizing and generalizing many such observations allows us to develop new \emph{laws and theories} that explain and predict how general concepts interact with one another holistically.
One well-known example of this is the theory of plate tectonics, which unified independent observations of fossil distributions matching across continents \cite{128463022}, seafloor magnetic striping \cite{4296143}, earthquake belt locations \cite{130687055}, and mid-ocean ridge spreading rates \cite{127076798} into a single coherent model of the Earth's geological dynamics.

Currently, the cost of theory development can be prohibitively high.
For example, building a theory around ``\emph{How AI explanations affect user trust}''\cite{277795373} requires gathering the hundreds of relevant papers, each reporting observations about similar and different concepts using varied phrasing (e.g., \emph{XAI} vs \emph{explainable AI}) or at different levels of abstraction (e.g., \emph{interactive} vs \emph{dialogue-based} explanations). %
Further, researchers need to reason at a \emph{useful level of abstraction}. A high-level statement such as ``\emph{system design affects user behavior}'' is commonly believed true but lacks utility; a specific statement such as ``\emph{Sonnet-3.7 outperforms GPT-4.5 when...}'' 
from a single paper may not generalize to scenarios in future studies. In contrast, aggregating comparisons between \emph{reasoning} vs \emph{non-reasoning models} from many papers, and across different tasks and contexts may yield more informative hypotheses about the nature of language models. 
Which level is useful, however, may depend on researchers' preferences, on the scope of a research project, on the alternative explanations being considered, and on what the evidence shows once aggregated at that level. Russell et al. found that schema revision to be the most costly step of sensemaking: a fixed schema is filled once, while a schema that is still changing is filled again at every revision \cite{207177544}.
Finally, beyond preferences, which concepts to focus on is sometimes not known in advance either. An experienced researcher may start from a few familiar ones, but must also explore the broader literature.
In this paper, we investigate how an interactive system can better support researcher-in-the-loop theory development grounded in broad literature, so that theory development can not only be a huge community-wide endeavor, but also a personal research ideation tool. %

We present \sysname{} (Figure~\ref{fig:system}), an interactive, literature-grounded theory development system. \sysname{} reduces hundreds of papers to a simple schema: each observation becomes a concept-relation triple (e.g., \triple{Nearest-Neighbor Explanation}{\edgeup{decreases}}{AI over-reliance}~\cite{255998499}), and concepts are organized into a shared ontology (e.g., \triple{Nearest-Neighbor Explanation}{\edgeup{is-a}}{Example-based Explanation}; Figure~\ref{fig:ontology}). Concepts in observations were extracted at the most specific abstraction level (e.g., a specific technique), while the ontology allows the user to aggregate the evidence at any level above them (e.g., a class of techniques). A researcher can therefore \emph{zoom} a concept up or down to see where a relation holds consistently and where the papers behind it disagree (Figure~\ref{fig:nesting}). For example, selecting \cpt{Nearest-Neighbor Explanation} allows a researcher to examine experimental observations about it, but zooming out to \cpt{Example-based Explanation} allows them to also consider all other types of explanation techniques of the same category. %

Combining a chat agent and a direct manipulation canvas (Figure~\ref{fig:system}), the researcher curates an \emph{evidence graph} of the concepts and relations they wish to focus on. This graph is an externalized mental model of the user and a \emph{shared representation}~\cite{277244228} between the user and the system. When the user curates the evidence graph, the agent is an exploration assistant. It searches the evidence from up to hundreds of papers and proposes additions anchored on what the researcher has placed on the canvas (Figure~\ref{fig:triples}). When the user switches focus to ideate hypotheses, the agent becomes a thinking partner. It generates a list of hypotheses\footnote{\label{fn:laws-vs-hypotheses}Prior work on generating such statements from literature, including \cite{284981954} whose approach we adapt, calls them ``laws.'' We use ``hypotheses'' instead: \emph{lawhood} is a status the scientific community confers over time, in the way Merton described eponymy as an honor the community bestows \cite{147253305}, and machine-generated statements have earned no such standing, so presenting them to users as laws would be overly strong.}  based on the evidence graph for the researcher to review and triage. The ones the researcher keeps are synthesized into a theory grounded in the evidence on their canvas (based on methods from \cite{284981954}.) Throughout, the researcher and the agent share the same vocabulary of concepts at different levels of the ontology, relations between the concepts, and the distinction between what a paper observed and what its authors claimed. The granularity of these structures and steps forms a ``sensemaking ladder''\cite{237007033}, where at each step the researcher decides what continues to the next.

\label{sec:intro-example}To illustrate further, in the same corpus, \cpt{Example-based Explanations} increases \cpt{Decision-Making Accuracy} in five papers, but it also increases both \cpt{AI Overreliance} and \cpt{AI Underreliance}. One hypothesis \sysname{} generated proposes a condition that reconciles all three (\S\ref{app:hyp-asymmetric}), in which examples move users from accepting a prediction to weighing evidence against it. Reliance therefore shifts both ways while accuracy still rises, because the gain where the AI errs outweighs the loss where it does not \cite{251224004}.
The hypothesis also included its scope and limitations. Specifically, it applies to teams where the AI errs often enough for users to catch (an accuracy of roughly 50 to 80\%), and its advantage diminishes as AI accuracy approaches 100\% and few errors remain to detect.%

Concretely, this paper makes the following contributions:

\begin{enumerate}
\item \sysname{}, a system that helps ideate new hypotheses and theories from hundreds of papers, based on 
an interaction design in which researchers work at the level of abstraction they choose: evidence and observations from hundreds of papers is aggregated along an ontology, so concepts can be zoomed, nested, and compared across levels;
\item The curated evidence graph as a \emph{shared representation} between researcher and system, which anchors the agent-assisted literature exploration and steers and grounds hypothesis generation;
\item A field deployment with 10 computer science researchers and two case studies with 2 immunology researchers and 2 agriculture researchers, which showed that researchers curated by their own preferences and project context, paying attention to the level of generality in concepts and hypotheses, and attended closely to conflicting evidence, while domain scientists found mechanisms outside their standard analyses and hypotheses potentially worth testing in new experiments.
\end{enumerate}

\section{Related Work}

\subsection{Interactive Literature Synthesis Tools}
LLMs are increasingly used for finding, reading, and synthesizing literature. In a survey of 816 publishing authors, 81\% reported having already incorporated them into their workflows, with discovering and synthesizing papers among the most common uses~\cite{273950276}. These activities are well studied in HCI: interactive systems have helped researchers make sense of networks of papers~\cite{6373651, 7511242}, collect and organize threads of related work~\cite{251402552, 260899915}, and build structured overviews of a research area~\cite{811620}. A growing class of ``deep research'' tools now serves this demand. Given a question, systems such as \cite{280529654} and commercial tools\footnote{\url{https://elicit.com} or \url{https://consensus.app/}} retrieve papers and compose them into organized reports.
However, these tools focus on summarizing and understanding existing literature; they do not take the additional step of generating new research ideas, hypotheses, or theories.

More closely related to our work are research ideation support tools. Most rely on the model's parametric knowledge, generating pools of candidate ideas~\cite{260467886}, research questions~\cite{269748457}, and support group ideation~\cite{267897885, 270638878}. More recent work also grounds ideas in literature. 
For example, IdeaSynth~\cite{273186404} grounds a researcher's idea in the literature and LitPivot~\cite{287121341} further help pivot it when relevant prior work was found.
However, these system only aim at improving proposal for a single future study, verified against relevant literature largely to ensure novelty. 

In contrast, this project set out to explore broad hypotheses and theory development, where a hypothesis may aggregates and extend hundreds of observations extracted from the literature.
Interactive tools for developing theories remain scarce. %
Langley~\cite{15548809} pointed this out in 2000, calling interactive interfaces for computational scientific discovery \emph{critically underexplored}; surprisingly, 25 years later, Velloso and Hornb{\ae}k~\cite{278062769} still found that while the HCI community values theories, it has produced few tools for building them, and proposed synthesizing causal relations between concepts as a potential fruitful avenue, which inspired a key design of our system. %

\subsection{Computational Scientific Discovery}
Borrowing the framework proposed by Langley \cite{268713625}, who synthesizes over four decades of research on computational scientific discovery originating with Simon \cite{178776253}, we frame theory building as producing three categories of scientific knowledge, each building on the one before. \emph{Taxonomies} identify the important concepts within a domain as a hierarchy of abstraction levels (e.g., the Linnaean biological taxonomy \cite{10075179}). \emph{Laws}\footref{fn:laws-vs-hypotheses} summarize many observations to describe relations between concepts. For example, Fitts's Law \cite{501599} predicts cursor movement time from the distance to a target and its size. Finally, \emph{Theories} are holistic models that explain a set of laws and their underlying observations. For example, Information Foraging Theory \cite{16487211,86675558} to explain browsing, link-following, and search under a single model (which was also inspired by the Optimal Foraging Theory \cite{86675558})

Recent work has made new progress on automating hypothesis and theory generation with LLMs\cite{248512482}. One thread focused on data. DataVoyager~\cite{267770682} calls for end-to-end data-driven discovery with LLMs, where hypotheses are searched and verified purely over provided tabular numeric datasets. AutoDiscovery~\cite{280150869} drops even the human-specified research question, exploring a dataset's hypothesis space guided by \emph{Bayesian surprise} and where experimental results most shift a model's beliefs. Both run without a researcher in the loop. In contrast, a design probe with 22 participants~\cite{277244228} found that an interactive tree diagram, a shared representation the user and the model both act on, put ``guardrails'' on AI-assisted hypothesis exploration, structuring the workflow and enabling backtracking, while AI-generated hints eased turning abstract ideas into data-backed ones, highlighting the benefits of putting researcher into the loop.

Most relevant to our work, Theorizer~\cite{284981954} generates theories from hundreds of papers. By holding generation to a fixed knowledge cutoff (i.e., using only papers published and models trained before a certain date), they show that the resulting theories can be novel and predictive of results published after the cutoff date. However, their pipeline is fully autonomous that takes a short query to a full theory, with no opportunity for researchers to give inputs to steer generation toward the concepts they wish to focus on. We build on this work, adapting its hypothesis and theory generation prompts, which are validated at scale but do not support researcher-in-the-loop, and extending it with an interactive interface that lets researchers explore the literature and steer theory generation based on their own preferences.

\section{Design Goals}

This project is inspired and informed by the decades of research described above. In particular, we are motivated by the insight succinctly articulated by Džeroski, Langley, and Todorovski in their book on computational scientific discovery~\cite{12808031}:

\begin{quote}
\emph{
Most [scientific] discovery research has focused on automated systems, yet few scientists want computers to replace them. Rather, they want computational tools that can assist them in constructing and revising their models. To this end, we need more work on interactive systems that let users play at least an equal role in the discovery process.}
\end{quote}

\noindent More specifically, we formulate four design goals, informed by both automated scientific discovery, scientific process of theory development~\cite{284981954,267770682,280150869} and a call-to-action in the HCI community~\cite{12808031}:

\begin{itemize}
  \item[\textbf{DG1}] {Extract experimental observations across many papers as concept-relation triples connected by an ontology.}
  \item[\textbf{DG2}] {Allow users to explore the literature for new insights by expanding from their prior knowledge, and curate an evidence graph with concepts at the level of abstraction they prefer.}
  \item[\textbf{DG3}] {Allow users to steer hypothesis and theory generation using the evidence graph as a shared mental representation.}
  \item[\textbf{DG4}] {Visualize the above interactively so users can ground generated theories in concrete observations from the literature.}
\end{itemize}

\begin{figure*}[t]
  \centering
  \includegraphics[width=0.9\linewidth]{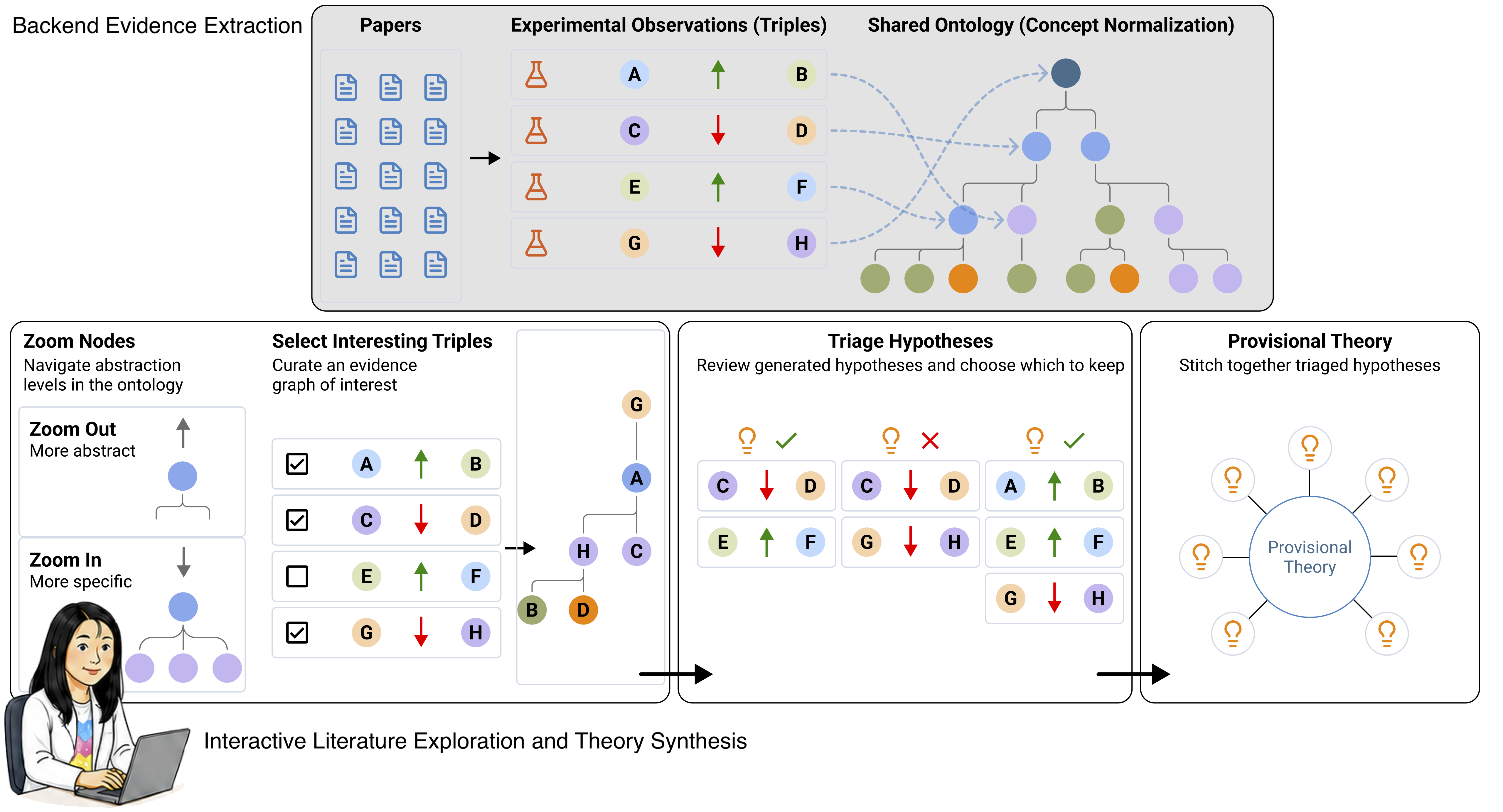}
  \caption{\sysname{} extracts experiments and observations as concept-relation triples from a large set of papers, and infers a shared ontology that normalizes and generalizes concepts across papers (top). The researcher can steer theory development by selecting concepts across abstraction levels and relations between them to curate an evidence graph (left). Based on this shared representation, \sysname{} generates hypotheses for her to review and select (center), and draws the hypotheses she keeps into a unified theory that stays grounded in her evidence graph (right).}
  \Description{A diagram of the \sysname{} process in four panels. The top panel shows a stack of paper icons with an arrow into a list of four extracted triples, each drawn as a flask icon, a lettered source concept, a green up arrow or a red down arrow, and a lettered target concept. Dashed lines run from each triple to the leaves of a tree on the right headed Shared Ontology, Concept Normalization, whose nodes are colored circles arranged over four levels. The lower left panel, outlined in green, is headed Zoom Nodes, Navigate abstraction levels in the ontology, and Select Interesting Triples, Curate an evidence graph of interest. It holds two controls reading Zoom out, more abstract, and Zoom in, more specific, a checklist of the same four triples with three of them ticked, and a laptop screen showing a node-link evidence graph over the eight lettered concepts. A researcher in a lab coat sits at a laptop beside it. An arrow leads down to a panel outlined in orange, headed Triage Hypotheses, Review generated hypotheses and choose which to keep, holding six cards; each card groups two or three triples under a lightbulb icon and is marked below with either a green tick or a red cross, four ticks and two crosses in total. An arrow leads right to a panel outlined in blue, headed Provisional Theory, Stitch together kept hypotheses, showing five lightbulbs on a ring around a central circle labelled Provisional theory. Below it a dashed arrow annotated outside scope.}
  \label{fig:overview}
\end{figure*}
\section{\sysname{}: User Scenario \& Synthesis Pipeline}
\label{sec:system}

We present \sysname{} (Figure~\ref{fig:system}), a system that scaffolds user-driven, literature-grounded theory development by allowing users to interactively progress from exploring existing ideas from the literature and hypothesizing new ones. Starting broad with a theory query, such as, \emph{Under what conditions does collaboration between humans and AI produce better decisions than either humans or AI working independently?} (from P6), and a set of 100--200 relevant papers, a user first explores and discovers key concepts and relations studied across multiple papers, curating an \emph{evidence graph} on their canvas (DG1, DG2). They then review and triage hypotheses generated from different regions of their evidence graphs, keeping the ones they find interesting (DG3). Finally, they can save the selected hypotheses by generating an overarching theory that connects them, linking the theory back to evidence on their graph, including evidence it leaves \emph{unaccounted for} or that \emph{conflicts with} its claims (DG4).

To illustrate how these parts work together, we describe how \persona{}, a researcher studying human-AI decision-making, works out how different types of AI explanations affect the performance of human-AI teams. %

\subsection{Building an Evidence Graph from Papers}

\begin{figure}[t]
  \centering
  \fbox{\includegraphics[width=\pagefrac{0.55}]{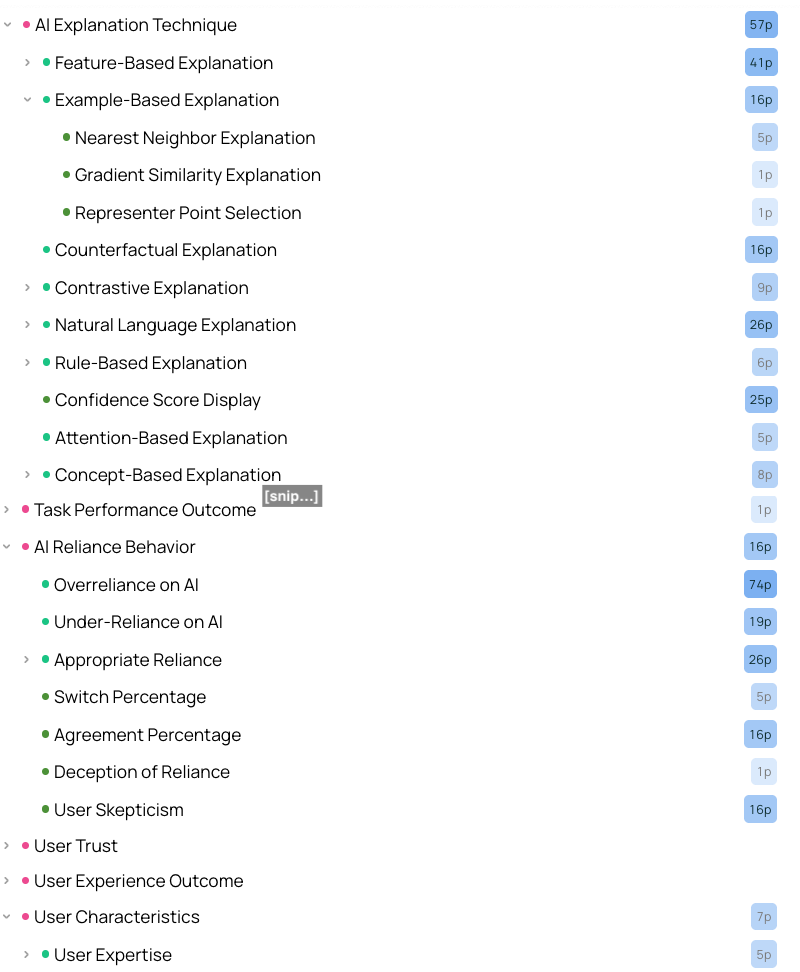}}
  \caption{Part of the ontology \sysname{} inferred for \persona{}'s corpus, shown in the outline panel. Each row carries the number of papers behind it: \emph{Overreliance on AI} mentioned in 74 papers while \emph{Deception of Reliance} only in 1. The ontology also powers the concept-zoom interaction, where zooming in on a concept offers the ones nested under it and zooming out merges it into its parent (Figures~\ref{fig:node-panel} and~\ref{fig:nesting}).}
  \Description{A screenshot of a collapsible outline panel listing ontology concepts as an indented tree. Each row has a coloured dot, a concept name, and a badge on the right giving a paper count. Top-level rows with pink dots read AI Explanation Technique with 57 papers, Task Performance Outcome, AI Reliance Behavior, User Trust, User Experience Outcome, and User Characteristics. Indented under AI Explanation Technique, rows with green dots read Feature-Based Explanation with 41 papers, Example-Based Explanation with 16, Counterfactual Explanation, Contrastive Explanation, Natural Language Explanation, Rule-Based Explanation, Confidence Score Display, Attention-Based Explanation, and Concept-Based Explanation. A third level under Example-Based Explanation lists Nearest Neighbor Explanation, Gradient Similarity Explanation, and Representer Point Selection. Under AI Reliance Behavior the rows include Overreliance on AI with 74 papers, Under-Reliance on AI with 19, and Appropriate Reliance with 26.}
  \label{fig:ontology}
\end{figure}

\persona{} has already read a few papers on the topic, but existing works test one or two explanation techniques at a time on different tasks, using inconsistent terminology and arriving at seemingly conflicting results.
Her aim is to develop unifying theories, but faced with the fragmented resources and the limited set of papers she happens to have read, Nora worries that her exploration might be skewed. Yet, she realizes that checking the picture in her head against the wider literature would take weeks or even months.
This leads Nora to turn to \sysname{}. To start, she uses a freely available paper search engine to search for ``papers that explore different types of explanations and their effects on human-AI teams'' and imports around 100 papers titles into \sysname{}. Using the Semantic Scholar APIs\footnote{\url{https://www.semanticscholar.org/product/api}} to retrieve the papers, \sysname{} extracts from each paper its experiments and observations as concept-relation triples (e.g., \triple{LIME explanations}{\edgeup{increases}}{AI overreliance}\cite{255998499}), yielding roughly 1,700 relations over 500 concepts. It also infers an ontology (Figure~\ref{fig:ontology}) that normalizes the same concept across papers and links specific concepts to broader ones they can be generalized to.

\subsection{User-driven Concept-relations Discovery}

\persona{}'s 100 papers covers a wide range of observations, such as task accuracy and completion speed, on users' cognitive load, or on how explanation effects shift with users' domain expertise. Which direction is worth developing depends on both the researcher's interests and how much evidence is available. Prior work on literature-grounded theory generation leaves this choice to the language model, which generates theories from a retrieved corpus fully autonomously~\cite{284981954}. In \sysname{}, the choice belongs to the researcher. The canvas is an externalized mental model of which concepts and relations her hypotheses should be built on (DG2, DG3). To support broad exploration and discovery, \sysname{} provides several complementary affordances described below. %

\subsubsection{Browsing all triples.}
\label{sec:all-triples}
\persona{} first skims the full list of triples, ranked so that the best-supported ones come first. When \persona{} clicks on a triple, the nodes and edge appear on her canvas. One challenge here is that the same relation can be listed at more than one level of abstraction. The papers behind \triple{LIME}{\edgeup{increases}}{AI Overreliance} are also evidence for the same relation between the broader concepts above \cpt{LIME}, and listing all of them would fill the list with restatements of a single finding. To address this, \sysname{} scores each relation at each level of the ontology, preferring the ones supported by more papers that agree with each other, and penalizing concepts broad enough to hide distinct relations beneath them. When several levels of the same relation compete, \sysname{} lists only the highest scoring one (\S\ref{app:triple-ranking}). %

\begin{figure}[t]
  \centering
  \fbox{\includegraphics[width=\pagefrac{0.55}]{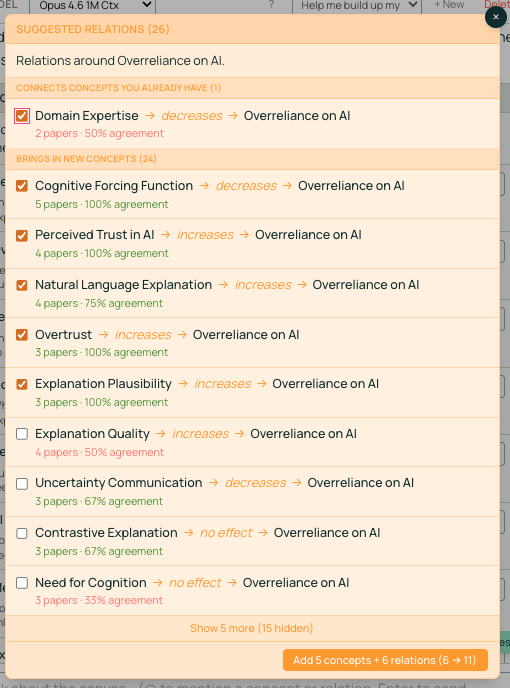}}
  \caption{The triage widget the agent returns instead of writing to the canvas itself, here for the 26 relations it found around \emph{Overreliance on AI}. Rows are split by what they would cost: the first group connects concepts already on the canvas, the rest bring in new ones. Each row carries the relation, the number of papers behind it, and how much those papers agree, and the best-supported rows arrive pre-checked.}
  \Description{A screenshot of a modal dialog inside the chat panel, headed SUGGESTED RELATIONS with a count of 26 and the line: Relations around Overreliance on AI. Rows are grouped under two orange headers, CONNECTS CONCEPTS YOU ALREADY HAVE with one row, and BRINGS IN NEW CONCEPTS with twenty-four. Each row has a checkbox, a source concept, an italic relation such as increases, decreases, or no effect, the target Overreliance on AI, and a small line reading a paper count and an agreement percentage. The first six rows are checked and the rest are not. A link at the bottom reads Show 5 more, 15 hidden, above an orange button reading: Add 5 concepts plus 6 relations, 6 to 11.}
  \label{fig:triples}
\end{figure}

\subsubsection{Expanding from a selected node.}
\label{sec:expanding-from-node}
\persona{} can click to select a concept node on her canvas to explore concepts around it in the right side panel in three ways. First, the panel lists relations to concepts not yet on the canvas. For example, selecting \cpt{LIME} shows its positive and negative downstream effects around \cpt{reliance} and \cpt{decision accuracy}.
Second, the panel lists the concept's ontological neighbors. She learns that the ontology groups \cpt{LIME} with other AI explanation techniques she knows less well, and adds one of them, \cpt{SHAP} \cite{21889700}. The third way is through hypotheses. The panel surfaces pre-generated hypotheses that involve the selected concept but also cite additional triples not yet on her canvas. One hypothesis, titled \emph{Expertise moderating the benefit of explanations}.
She marks the hypothesis as \emph{interesting}, which expands her canvas to include all the triples it references and saves the hypothesis into a triage drawer (\S\ref{sec:triaging-hypothesis}).

\subsubsection{Asking the chat agent.}
\label{sec:asking-agent} 
Beyond direct manipulation on the canvas, \persona{} can also converse with the chat agent to more flexibly explore the literature. The chat agent has the the current evidence graph she has curated in the context, but also a set of tool to explore the full list of triples and full text of the papers. \persona{} can ask about concepts she knows as well as about ones she does not know to look for. She starts by naming techniques she wants to cover, such as \cpt{Example-based Explanations} and \cpt{Confidence Score Display}, and later asks the opposite: \emph{which well-studied themes does my evidence graph not yet cover?} In both scenarios, the agent answers with a card of concepts not yet on her canvas (Figure~\ref{fig:system}, left). Clicking a concept places it alone on the canvas and opens a dialog of the relations around it (Figure~\ref{fig:triples}). The best-supported relations are pre-checked. Around \cpt{Confidence Score Display}, \persona{} notices a well-supported link to \cpt{AI Overreliance} she did not know of and keeps it, unchecks one row she considers off-topic, and adds the rest as a batch. \cpt{LIME explanations}\cite{13029170} enters her canvas through the second kind of question. She has heard of the technique but is not planning to include it, until the card shows how heavily it has been studied. Finally, when her canvas has split into two disconnected clusters, she asks the agent to check if there are any concepts from the literature that connects them. The agent proposes a pathway through a concept she does not yet have, \cpt{Example-based Explanations}\edgeup{increases}\cpt{AI Underreliance}\edgedn{decreases}\cpt{Human-AI Team Performance}\cite{53039904,258187500}.

\begin{figure}[t]
  \centering
  \fbox{\includegraphics[width=\pagefrac{0.55}]{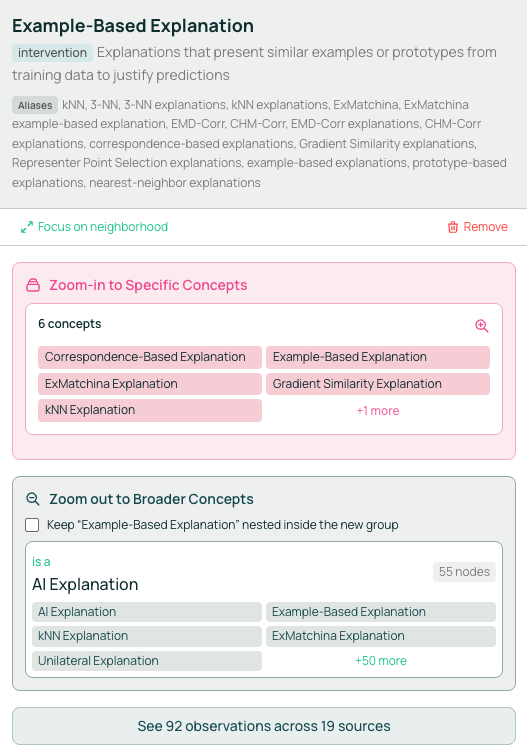}}
  \caption{The concept panel for \emph{Example-Based Explanation}. The two controls below move the concept along the ontology. Zooming in replaces it with the six specific techniques grouped under it; zooming out replaces it with its parent, \emph{AI Explanation}, and the checkbox keeps the current concept nested inside that parent instead of folding it away, which is the state Figure~\ref{fig:nesting} shows on the canvas.}
  \Description{A vertical side panel. At the top, the concept name Example-Based Explanation, a category chip reading intervention, a one-sentence definition, and a long list of aliases. Below are two boxed sections. The first, Zoom-in to Specific Concepts, lists six concepts including Correspondence-Based Explanation, ExMatchina Explanation, and kNN Explanation. The second, Zoom out to Broader Concepts, shows the parent AI Explanation with a badge reading 55 nodes, a checkbox for keeping the current concept nested inside the new group, and a sample of the concepts in that group. A footer button reads: See 92 observations across 19 sources.}
  \label{fig:node-panel}
\end{figure}

\subsubsection{Zooming and nesting.}
\label{sec:zooming-nesting}
\begin{figure}[t]
  \centering
  \includegraphics[width=\pagefrac{0.55}]{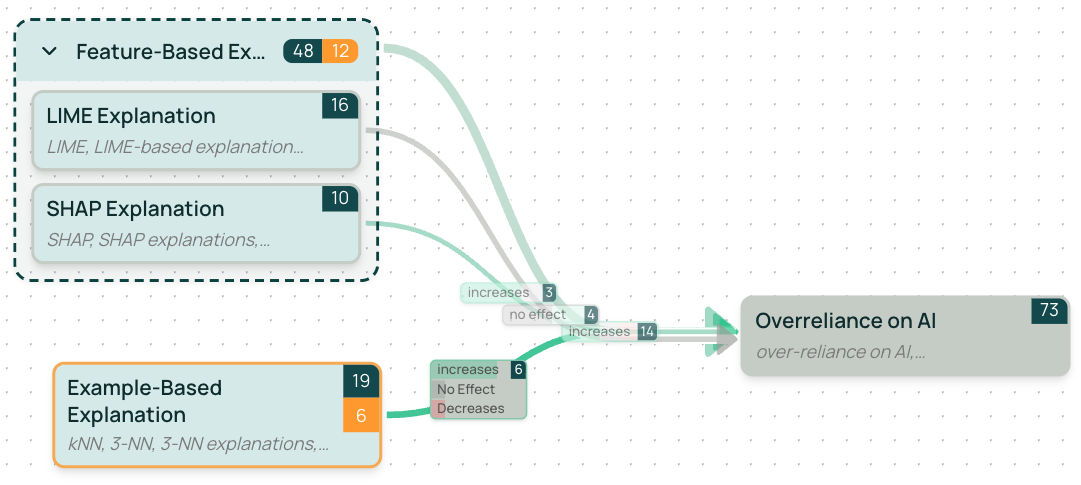}
  \caption{Nesting keeps a concept's aggregated evidence without hiding the techniques beneath it. \emph{Feature-Based Explanation} is drawn as a group with \emph{LIME} and \emph{SHAP} nested inside it, so the group's merged edge to \emph{Overreliance on AI} and the two techniques' own edges are on the canvas at the same time. Figure~\ref{fig:node-panel} shows the controls that produce this state. On a node, the dark pill counts the unique papers behind the concept and the orange pill the number of sub-concepts folded into it. Hovering over the edge out of \emph{Example-Based Explanation} shows the detailed distribution of reported relations from different papers. Clicking on the edge leads to Figure~\ref{fig:evidence}.}
  \Description{A screenshot of a small region of the canvas. On the left, a group node drawn with a dashed border is labelled Feature-Based Explanation and contains two child boxes, LIME Explanation and SHAP Explanation, each with its own source-count badge. Curved edges leave the group and each child and converge on a node labelled Overreliance on AI. Their labels read increases with 3 sources, no effect with 4 sources, and increases with 14 sources. Below, a separate node labelled Example-Based Explanation connects to the same target through an edge whose label is a stack of three rows reading increases, No Effect, and Decreases.}
  \label{fig:nesting}
\end{figure}

\persona{} notices that \cpt{LIME} and \cpt{SHAP} are very specific techniques each carrying only few observations on her canvas. From the node panel she zooms out to their shared ontology parent, \cpt{Feature-attribution Explanations} (Figure~\ref{fig:node-panel}). The canvas re-aggregates, merging the two techniques' edges and roughly doubling the evidence behind each relation (DG2). This allows her to build coherent theories comparing and contrasting \cpt{Feature-attribution Explanations} against \cpt{Example-based Explanations}. Zooming out gains evidence but loses the distinction between the two techniques, so she can instead nest them inside their parent as a group, which keeps the aggregated relations while the individual techniques remain visible within it (Figure~\ref{fig:nesting}).

\subsection{Drilling into the Evidence}

By this point the canvas holds the concepts \persona{} cares about, and before building on them she wants to know how much to trust them. One way \sysname{} convey uncertainty is to visualize the amount of evidence behind concepts and relations, and how much they agree with one another. Each nodes shows how many unique papers reported observations about it (in teal) and how many child concepts are grouped (in orange), allowing her to see which concepts are heavily studied. Clicking a node shows the concept's definition, its aliases, and the more specific concepts the ontology groups beneath it (Figure~\ref{fig:node-panel}).

\begin{figure}[t]
  \centering
  \fbox{\includegraphics[width=\pagefrac{0.55}]{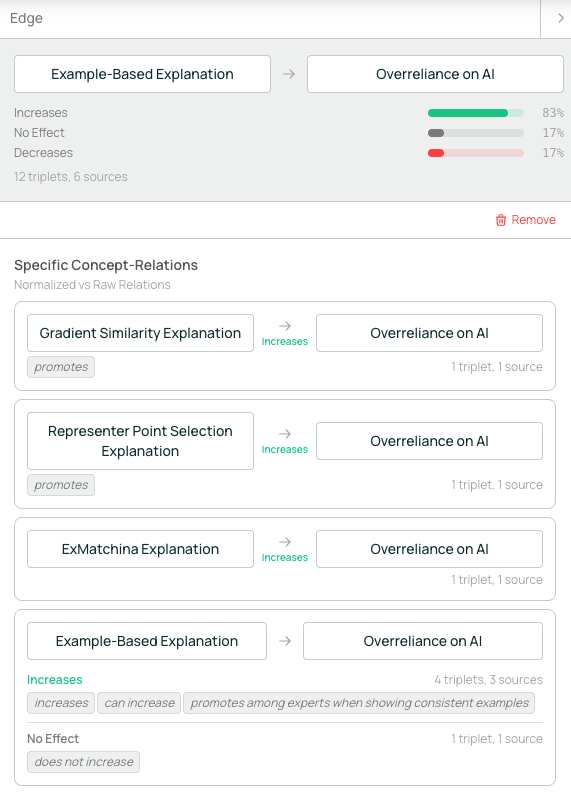}}
  \caption{The evidence panel for the aggregated edge from \emph{Example-Based Explanation} to \emph{Overreliance on AI}. The bars break the 12 triples behind the edge down by relation as a share of the 6 unique papers reporting them; the shares add to more than 100\% because a paper that reports two different relations counts toward both. Below, the panel unfolds the aggregate into the specific concept-relations it merged, each with the raw phrasing its paper used next to the canonical relation it was normalized to. Each of these unfolds one step further into the observation and experiment behind it (Figure~\ref{fig:reader}).}
  \Description{A vertical side panel headed Edge. At the top, two boxes joined by an arrow read Example-Based Explanation and Overreliance on AI. Beneath them, three labelled bars read Increases 83 percent, No Effect 17 percent, and Decreases 17 percent, followed by the line: 12 triplets, 6 sources. The lower part of the panel, headed Specific Concept-Relations, Normalized versus Raw Relations, is a stack of cards. Each card names a more specific source concept, such as Gradient Similarity Explanation or Representer Point Selection Explanation, with the same target, the normalized relation, the raw relation phrases in small chips, and a count of triplets and sources.}
  \label{fig:evidence}
\end{figure}

\begin{figure*}[t]
  \centering
  \fbox{\includegraphics[width=\linewidth]{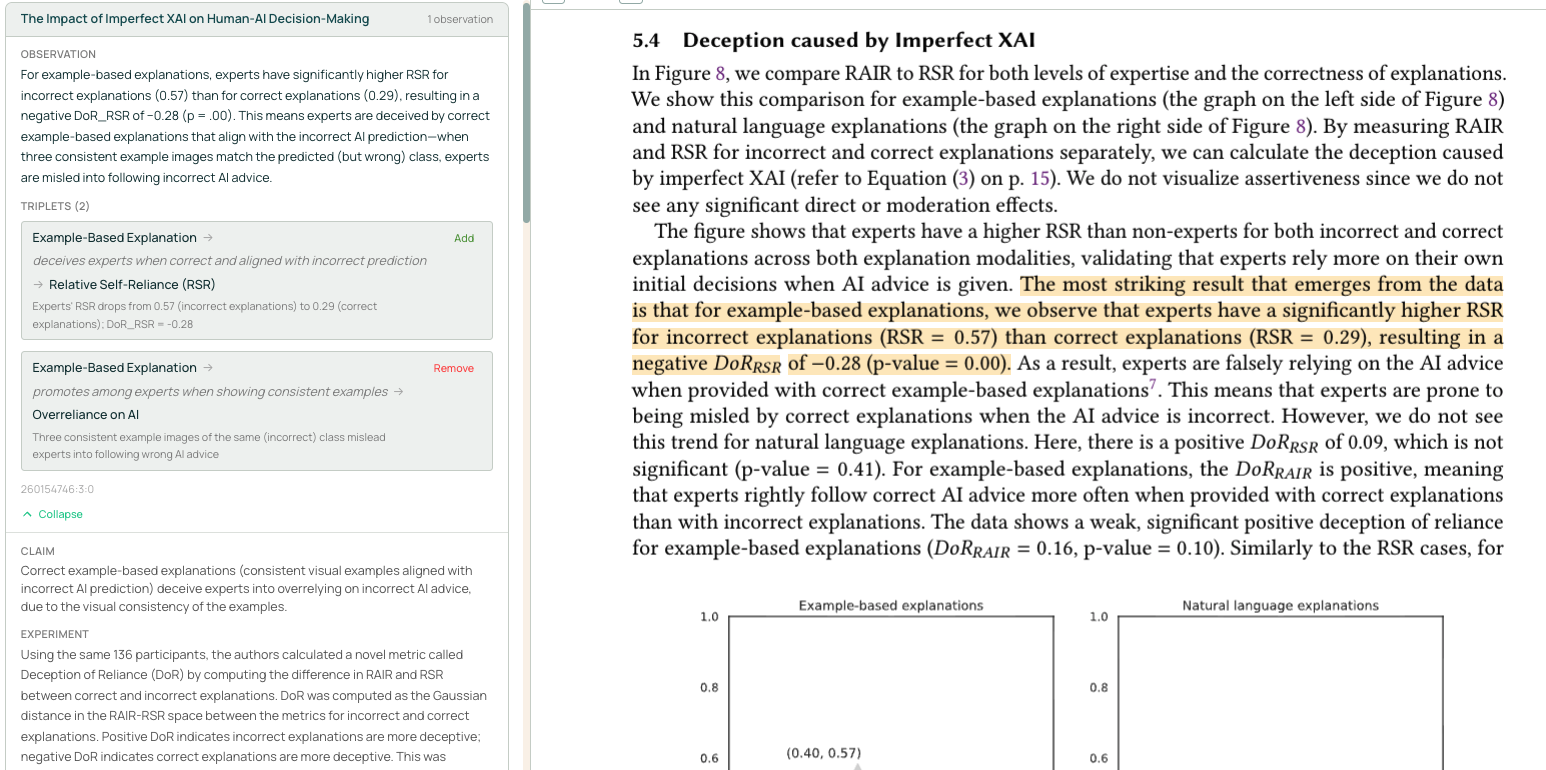}}
  \caption{Every relation on the canvas traces back to sentences in a paper. This is the last step of the drill-down that starts at an edge in Figure~\ref{fig:evidence}. \textbf{Left:} the evidence behind one triple, showing the observation as it was extracted, the triples drawn from that observation, the claim the authors made from it, and the experiment that produced it. \textbf{Right:} the same paper opened in the reader, with the sentence the observation was extracted from highlighted in place, so a user can check the extraction against the authors' own wording and read around it. (Example paper \cite{260154746}).}
  \Description{A two-pane screenshot. The left pane is a scrolling panel headed with a paper title, The Impact of Imperfect XAI on Human-AI Decision-Making, and the note 1 observation. It contains an OBSERVATION paragraph reporting relative self-reliance figures for expert participants, a TRIPLETS section with two extracted concept-relation triples each carrying an Add or Remove button, and CLAIM and EXPERIMENT sections describing what the authors concluded and how the study was run. The right pane shows the rendered PDF of that paper, open at a section headed Deception caused by Imperfect XAI, with one long sentence highlighted in yellow, the same sentence the observation was extracted from.}
  \label{fig:reader}
\end{figure*}

Clicking a node shows the concept's definition, its aliases, and the more specific concepts the ontology groups beneath it (Figure~\ref{fig:node-panel}). 
For example, the edge between \cpt{Example-based Explanations} and \cpt{AI Overreliance} is supported by 6 papers, 5 of which reported an increase. Clicking an edge opens an evidence panel (Figure~\ref{fig:evidence}) that unfolds the aggregate into the specific concept-relations it has merged, each with the phrasing its paper used alongside the canonical relation it is normalized to, and from there into each source paper's extracted experimental setup and observation (Figure~\ref{fig:reader}). \persona{} drills into that edge, skims the evidence behind it, and reads the supporting passages of two papers directly. She comes away more confident that the aggregated increase reflects the studies themselves and not an artifact of extraction. Every concept and relation in \sysname{} traces back to specific sentences in specific papers (DG4). The chat agent also has access to the same extracted evidence and the full text of each paper, which allows \persona{} to also ask questions about the literature flexibly.

\subsection{Developing Hypotheses and Theories}

After she is happy with all the concepts and relations she has captured on her canvas, \persona{} switches focus from exploring the literature (DG2) to ideating hypotheses and theories (DG3).

\subsubsection{Generating and triaging hypotheses.}
\label{sec:triaging-hypothesis}
She clicks ``Generate hypotheses'' (based on my canvas) and further steers with a short prompt: \emph{predict how different explanation types affect user experience and task performance for users with different domain expertise}. \sysname{} streams in a batch of new hypotheses, each grounded in a different region of her canvas. She finds one particularly interesting, as it describes a pattern she has noticed in her own reading but never seen stated:
\begin{quote}
``\emph{feature-attribution explanations typically fail to improve decision-making accuracy and primarily function as persuasive signals that increase overreliance on incorrect AI outputs, whereas example-based explanations more reliably enhance accuracy by facilitating the detection of AI unreliability.}''
\end{quote}
\persona{} marks it as interesting, and continues to review and mark the rest of the hypotheses in the drawer.

\begin{figure*}[t]
  \centering
  \fbox{\includegraphics[width=\linewidth]{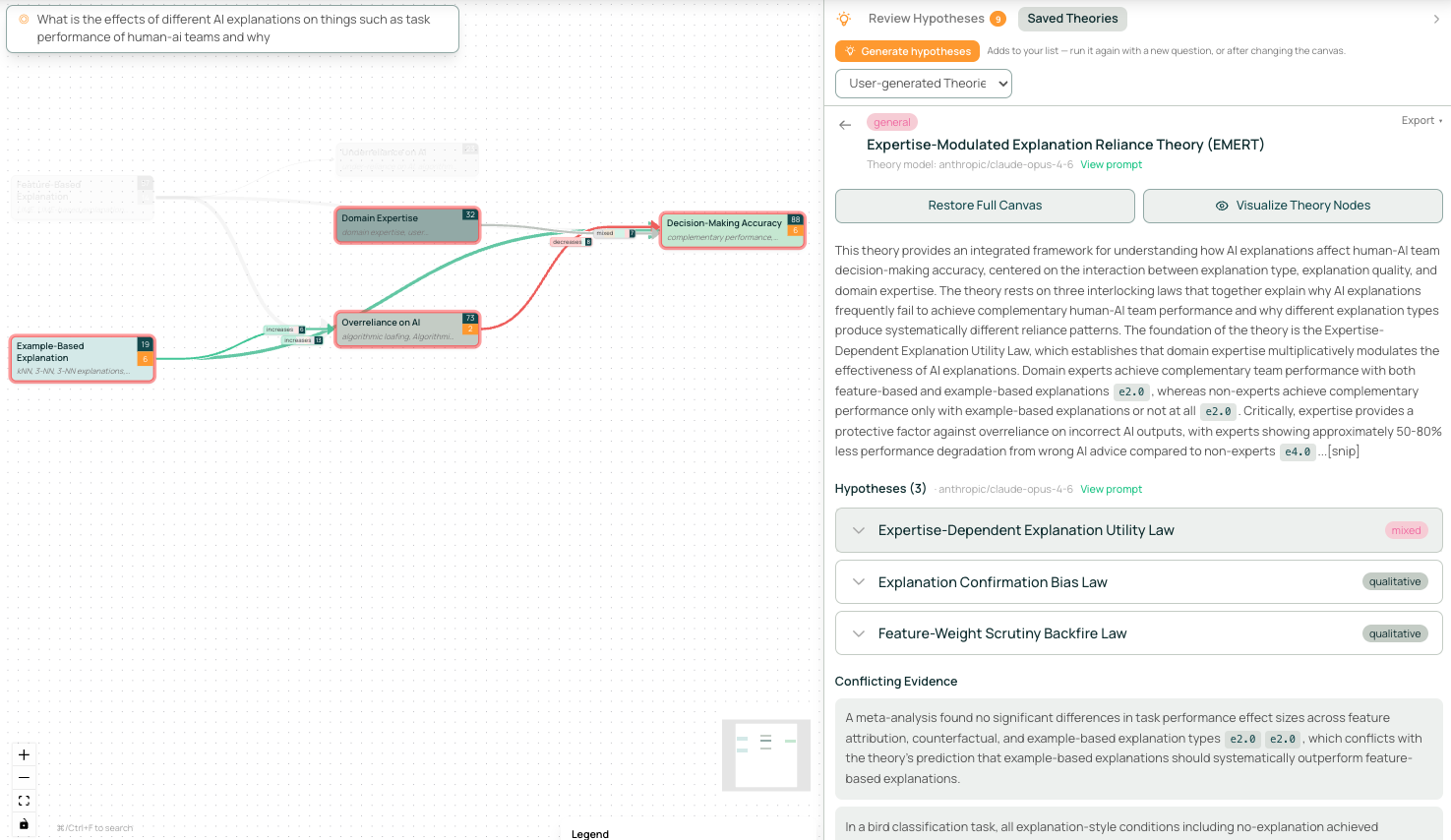}}
  \caption{A saved theory, synthesized from hypotheses kept in the review drawer on the right of Figure~\ref{fig:system}. The theory starts with a description that ties the hypotheses together under a single lens.
  The inline badges in the description reference the evidence behind each claim in the user-curated graph on the left, and hovering one highlights the corresponding edge.
  The hypotheses listed below can also be hovered to show the parts of the evidence graph they reference. The \emph{Conflicting Evidence} and \emph{Unaccounted For} sections report evidence on the canvas that runs against the theory or is not covered by it (based on \cite{284981954}).}
  \Description{A two-pane screenshot. The right pane is the theory panel. It shows the theory name, Expertise-Modulated Explanation Reliance Theory, a paragraph describing the three interlocking claims it is built from, with small inline badges labelled e2.0 and e4.0 embedded in the text, and buttons reading Restore Full Canvas and Visualize Theory Nodes. Below that, a section headed Hypotheses lists three collapsed rows, each with a name and a tag reading mixed or qualitative. At the bottom, a section headed Conflicting Evidence contains two grey boxes describing findings that run against the theory. The left pane is the canvas, where five concept nodes joined by green and red edges are drawn at full contrast while the remaining nodes are faded out.}
  \label{fig:theory-footprint}
\end{figure*}

\subsubsection{Assembling hypotheses into a theory.}
\persona{} selects five hypotheses and clicks ``Save as theory'' (DG3). \sysname{} assembles them, kept verbatim, into a theory based on \cite{284981954}: an overview statement that summarizes the saved hypotheses together, as well as listing evidence from her graph that is \emph{unaccounted for} or \emph{conflicts with} the theory. Hovering over each constituent hypothesis highlights the relevant concepts and relations (Figure~\ref{fig:theory-footprint}), and she notices that one hypothesis is supported by far fewer observations than the others and flags it as the natural candidate for her lab's next experiment judging it as underexplored by the literature (DG4).

\subsection{Backend Synthesis Pipeline}
To support the above user interaction, \sysname{}'s backend consists of the following six phases. The first four turn a corpus of papers into a graph; the last two generate hypotheses and then synthesize the selected ones into a theory. We fetch metadata and open-access full text from the freely available Semantic Scholar API\footnote{\url{https://www.semanticscholar.org/product/api}} and use Claude Opus 4.6 given the difficulty of the task. Base graphs are precomputed before interaction, and take about 30--60 minutes and cost 50--150 USD for processing around 100 papers.

\subsubsection{Phase 1: Initial Extraction.}
\label{sec:impl-extraction}
To separate observations (what happened during an experiment) and claims (generalizations made by the authors), we extract from each paper using the following information structure (\S\ref{app:prompt-findings}): a list of experiments reported and their designs, then the observations made during each experiment in both prose and triple forms, then the claims made supported by each experiment. In addition to the structured representation, we also keep long paragraph snippets with bounding boxes as evidence, so users can verify each extract with highlights in the papers' PDFs. The snippets also serve as additional evidence to consider for the subsequent hypothesis generation component.

\subsubsection{Phase 2: Ontology Generation.}
\label{sec:impl-ontology}
The second phase generates an ontology from the extracted experiment designs, observations, and claims (\S\ref{app:prompt-ontology}). 
Here, this context is designed to provide signals for language models to capture how researchers in the community abstract their observations into broader claims. For example, when papers test individual AI explanation techniques, such as \cpt{LIME} or \cpt{SHAP}, the authors often discuss their observations using more general terms, such as \cpt{Feature-attribution Explanations} (Figure~\ref{fig:ontology}). The generated ontology is a list of concepts each with an ID, parent ID, short name, and a full definition.

\subsubsection{Phase 3: Re-extraction and Linking to Ontology.}
\label{sec:impl-linking}
The third phase repeats phase 1, but this time with the ontology in the input context (\S\ref{app:prompt-findings}) and additionally constrains the language model to use ontology concepts for the triples instead of free text. This allows  one paper's \cpt{algorithm transparency} and another's \cpt{model interpretability} to resolve to the same concept, so that the system can aggregate observations around the same concepts across papers. %

\subsubsection{Phase 4: Relation Normalization.}
\label{sec:impl-relnorm}
A short phase collapses the many distinct relations (e.g., ``significantly improves,'' ``enhances,'' ``prohibits'') in the triples onto at most twelve canonical labels, seven of them fixed and five left open to capture query-specific relations (\S\ref{app:prompt-relnorm}).\footnote{increases, decreases, no effect, correlates with, outperforms, underperforms, matches} %

\subsubsection{Phase 5: Hypothesis Generation.}
\label{sec:impl-hypothesis}
We adapted the hypothesis generation prompt from Jansen et al.~\cite{284981954} (\S\ref{app:prompt-theorizer}). In contrast to \cite{284981954}, which generates a fixed extraction schema for all papers and organizes evidence as a paper-facet table, we serialize the evidence graph as a list of evidence edges, each listing its two concepts, the relation distribution, an ID, and per-paper entries holding that paper's raw relation phrasing, claim, observation, and experimental design from Phase 1. Because these edges are the aggregated edges on the canvas, hypotheses are generated at the level of abstraction the user chose, while the per-paper entries beneath each edge let the model state conditions at the level of individual studies; nested concepts are serialized as nested objects so the model can take into account how the user organized them.

The pre-generated hypotheses that support the interaction described in \S\ref{sec:expanding-from-node} come from this same component. Since the full evidence graph often runs past the model's 1M context window, we sub-sample the edges over six runs, where each run samples a subset of edges weighted by the number of unique papers behind it and discounting that weight steeply once generated hypotheses have cited it (\S\ref{app:evidence-sampling}). %

\subsubsection{Phase 6: Theory Synthesis.}
\label{sec:impl-theory}
The final phase turns selected hypotheses into a theory, again adapting the prompt from Jansen et al.~\cite{284981954} (\S\ref{app:prompt-theorizer}). The user-selected hypotheses are included verbatim, plus an overarching statement tying them together under a single lens. 
It also considers the whole evidence set behind the canvas, and not only what the hypotheses already cite, so that the theory includes which of that evidence it leaves unaccounted for and which conflicts with its hypotheses.

In both Phases 5 and 6, the model references edges as inline XML tags to ground its hypothesis and theory descriptions. In the interface they show up as references, allowing users to hover any piece of generation and see the relevant parts of their evidence graph highlighted. From there, they can drill-down all the way to see evidence snippets highlighted in the PDF (Figure~\ref{fig:theory-footprint}).

\subsubsection{The Chat Agent.}
\label{sec:impl-agent}
Before each of the user's messages to the chat agent, we prepend a JSON snapshot of their canvas: every visible concept and relation with its paper counts and agreement, the full text of the findings behind whatever they have selected, the top-ranked addable triples adjacent to their concepts, the pool relations between concepts they already have but have not drawn, each concept's ontology ancestors and siblings and children, the papers in scope, and the goal they are working toward. %
The agent has eleven tools (\S\ref{app:tools}) allowing it load additional context based on the user's request, such as full text of papers, generated hypotheses, concepts and relations on their canvases and beyond, which allows users to ask questions not just about their canvas and hypotheses, but also help explore the full corpus.

\section{Generation Pipeline Technical Evaluation}
\label{sec:tech-eval}

The interactions described in \S\ref{sec:system} rely on the ontology to aggregate evidence around similar concepts across different levels of abstraction. We evaluated against the closest reference we found, the hierarchical outlines of survey papers, where an author's grouping of cited papers under one concept, typically organizes papers thematically. We drew these from a collection of 156 outlines extracted from computer science surveys, curated by Zhu et al.~\cite{281496866}, and sampled 12 whose referenced papers were at least 90\% open access so our pipeline could process their full text (spanning 12--48 concepts over 20--173 papers.)

It is noteworthy that survey outlines typically organizes what an entire paper is \emph{about} thematically, while our ontology for theory building has to support more specific concepts such as \emph{intervention} or \emph{measure}. For example, to express \triple{LIME}{\edgeup{increases}}{AI over-reliance} it must contain \cpt{AI over-reliance}, which papers measure but are rarely about, and which seldom appears in a survey outline. We therefore ask whether our ontologies \emph{recall} the survey's concepts, one concept at a time and at any level of the hierarchy, instead of asking whether they reproduce the same outline.

Zhu et al. propose a method that 
recovers a single outline matching the expert one, and it is built to replicate the survey authors' curation judgment, so its outlines stay close to the expert's in size: 15.8 concepts on average against 24.0 in the outlines themselves. Ours over-generates relevant concepts for interactive curation and indexes concepts beyond broad themes, at 167.5 on average. Because the two pipelines are rewarded for different properties, the comparison below is a reference point and not a claim that one method outperforms the other.\footnote{As a sanity check, we ran their implementation on the 12 survey outlines using the same models in our system and it scores within 0.04 of their published results (\S\ref{app:sanity-check})}

\begin{table*}[t]
  \caption{Per-concept best-match Jaccard over the 288 non-trivial survey concepts, ours against the reference method of Zhu et al.~\cite{281496866}. The first row pools all 288 concepts; the second computes each statistic within each ontology and averages across the 12 ontologies (95\% CIs over those 12 values).}
  \label{tab:per-concept-overlap}
  \footnotesize
  \centering
  \begin{tabular}{lrrrrr}
    \toprule
    Scope & Ours closer & Zhu et al. closer & Tie & Mean best-J (ours) & Mean best-J (Zhu et al.) \\
    \midrule
    Pooled 288 concepts & 204 (71\%) & 55 (19\%) & 29 (10\%) & 0.508 [0.484, 0.532] & 0.375 [0.354, 0.397] \\
    Averaged 12 ontologies & 74.0\% [64.3, 83.7] & 16.7\% [8.2, 25.3] & 9.3\% [3.1, 15.4] & 0.542 [0.477, 0.608] & 0.391 [0.350, 0.431] \\
    \bottomrule
  \end{tabular}
\end{table*}

\label{sec:per-concept}

For each survey concept at each level of its hierarchy, we found the concept in each pipeline's output with the highest paper-set overlap using Jaccard similarity (shared papers over combined unique papers) and compared which pipeline came closer. We excluded 51 survey concepts that either covered all or a single paper which could inflate our score. Over the remaining 288, our ontologies contained the closer match for 204 (71\%), Zhu et al.'s method for 55 (19\%), with 29 ties (Table~\ref{tab:per-concept-overlap}). Because larger survey ontologies contribute more concepts to these pooled counts, we also computed the win rate within each ontology; it averaged 74.0\% across the 12 ontologies and exceeded 50\% on each of them. Mean best-match Jaccard was 0.508 for our ontologies against 0.375 for Zhu et al. (paired $t(287){=}11.0$, $p<0.001$). 

To see where the two pipelines diverge, we examined the matches themselves. Our pipeline wins when its ontology names the survey's concept almost verbatim and links the same papers to it:

\begin{itemize}
    \item \textbf{Tasks $\rightarrow$ Geometry Problem Solving} (4p)~\cite{254877175}:\\\texttt{Math. Reasoning Task $\rightarrow$ Geometry Problem Solving} (Ours. 4p. $J{=}1.00$.)\\``Geometry Problem Solving Combining Textual and Diagrammatic Inputs'' (Zhu et al. 3p. $J{=}0.75$)
\end{itemize}

\noindent Losses arise where the paper sets diverge, because our pipeline links each paper to the concepts its observations emphasize, while survey authors group papers by editorial judgment of what each paper is \emph{about} as a whole:

\begin{itemize}
    \item \textbf{Pre-trained Models $\rightarrow$ Multilingual $\rightarrow$ Early Exit} (7p)~\cite{212747830}:\\\texttt{Model Compression Method $\rightarrow$ Early Exit / Adaptive Inference} (Ours. 5p. $J{=}0.71$.)\\``Early Exit Mechanisms for Efficient Inference in Pre-trained Language Models'' (Zhu et al. 7p. $J{=}1.00$)
\end{itemize}

\noindent Finally, most of our generated concepts are not the closest match for any survey concept. However, they capture concepts useful for theory-building that are rarely a theme survey papers. For example, we generated \cpt{Hallucination} in a math language models survey~\cite{266191124}, where papers typically report it as a limitation and not a main theme of the papers, but users of our system may wish to develop hypothesis around hallucination. Whether researchers could find their way to them in practice is a question we return to in the field study (\S\ref{sec:field-study}).

\begin{table*}[t]
  \caption{The theory questions each participant worked on. Some created additional graphs during the study (Table~\ref{tab:study-engagement}). A session ends when there was 15+ minutes of inactivity.
   The final four rows are the two case-study teams (\S\ref{sec:case-studies}), whose engagement ran through meetings and self-directed use outside the deployment protocol, so minutes and sessions are not shown.}
  \label{tab:study-participants}
  \scriptsize
  \setlength{\tabcolsep}{2pt}
  \centering
  \begin{tabularx}{\linewidth}{@{}lrr>{\raggedright\arraybackslash}X@{}}
\toprule
\textbf{ID} & \textbf{Min.} & \textbf{Ses.} & \textbf{Theory question} \\
\midrule
P0 & 123 & 10 & How does the presence of external members (AI agents, new hires, new friends) affect group dynamics? \\
P1 & 109 & 10 & How do different context-management strategies in agentic AI systems affect task performance? \mbox{(and 2 more)} \\
P5 & 53 & 7 & How does different AI explanation technique affect users on cognitive load and task performance? \mbox{(and 1 more)} \\
P6 & 179 & 16 & How do different grounding strategies (retrieval, tool, human feedback) affect the novelty and rigor of AI-generated scientific discoveries? \mbox{(and 3 more)} \\
P7 & 90 & 5 & How do different AI interaction interface designs (conversational, embedded, agentic) affect user adoption, engagement, and task performance? \mbox{(and 1 more)} \\
P8 & 76 & 8 & How do different presentations of AI diagnostic advice [...] affect clinician overreliance and diagnostic accuracy in clinical decision support settings? \mbox{(and 1 more)} \\
P13 & 207 & 12 & How does pre-visit information influence visitor perceptions and expectations in cultural tourism? \\
P15 & 111 & 6 & How do different assistive technologies affect how low-vision people perceive and experience outdoor environments? \mbox{(and 1 more)} \\
P18 & 63 & 8 & How does Socratic questioning and guided discovery in AI tutoring affect learners' skill development? \\
P19 & 168 & 13 & How does query ambiguity and underspecification affect the behavior and task success of LLM-based research and search agents? \mbox{(and 1 more)} \\
\midrule
A1 & \multicolumn{2}{c}{2 month} & How the non canonical splicing process and the synthesis of terpenoids let coffee and maybe other plants to resist elevated air temperatures [...] \mbox{(and 2 more)} \\
A2 & \multicolumn{2}{c}{case study} & How spatial proximity to regional landrace richness hotspots affects the probability of re-emergence of locally lost landraces [...] \mbox{(and 1 more)} \\
\midrule
I1 & \multicolumn{2}{c}{2 month} & CD4 effector-memory T cells showing off-target transcriptional differences between ruxolitinib (free base) and ruxolitinib phosphate [...] \mbox{(and 2 more)} \\
I2 & \multicolumn{2}{c}{case study} & How adolescent (11--12) immune cells differ from adult cohorts in their response to influenza vaccination, per cell type \mbox{(and 1 more)} \\
\bottomrule
\end{tabularx}
\end{table*}

\section{Study 1: Field Deployment w/ CS Researchers}
\label{sec:field-study}

To investigate the costs and benefits of the interactive system, we conducted a field deployment study with 10 researchers, each using \sysname{} on knowledge graphs built from research questions they submitted, at their own pace, with at least one exported theory as the deliverable. %
\subsection{Participants}

We recruited participants form to Twitter and Slack research communities (\#openhci and \#opennlp with roughly 2,500 members across 100+ organizations). The form asked for demographic information and two to four theory questions the participant wanted relevant to their research., The following example queries were provided: ``How do different moderation policies and techniques affect the dynamics of online communities?'' This allowed us to pre-generate the graph before the study to save time.
Of the 21 researchers who signed up, 10 started and completed the study. All participants were graduate students in computer science and adjacent fields (6 male and 4 female; age: 23--37, M=27.3, SD=3.9), most of whom identified their area as HCI or NLP, AI, but also covers specific domains such as Computational Ecology, Algorithm, Computer Vision and Healthcare. %

\subsection{Procedure}

On-boarding and usage were unmonitored. Participants first watched a 12-minute tutorial video, then logged into the deployed system to find their graphs. Over 1--2 weeks, we asked them to spend one to two hours in the system, in as many sittings as they liked, and to produce one theory relevant to their research, email the theory as a PDF report to us to schedule an exit interview.

The exit interviews lasted 30 minutes and followed a semi-structured script over video call. Participants shared their screens and performed a retrospective walkthrough of their workspace: the canvases they built, the hypotheses they kept and dismissed, and the theory report they exported. Before each interview, we reviewed the participant's exported theory and interaction logs to prepare participant-specific questions, and in the interview we probed what felt useful, what did not, and what they would want next. Participants received a \$150 USD honorarium upon completing. This study was exempted by our institution legal team.

\subsection{Data Collection and Analysis}

Our analysis draws on three sources of data: interaction logs from the deployment, the artifacts participants produced, and the exit interviews. All interviews were recorded and transcribed. The first author analyzed the data using a thematic analysis method and went through the transcripts in multiple rounds of coding (Braun and Clarke~\cite{10075179,197748828}).%

\subsection{Findings}

\begin{table}[t]
  \caption{Engagement per participant, based on the interaction logs. Zoomed counts the child concepts revealed by zooming in or grouped by zooming out. Removed counts the distinct concepts a participant placed and later took off the canvas, by hand or by clearing it, and Canvases counts the canvases each participant saved.}
  \label{tab:study-engagement}
  \scriptsize
  
  \setlength{\tabcolsep}{3pt}
  \centering
  \begin{tabular}{lrrrrrrrrrrr}
  
\toprule
 & \textbf{P0} & \textbf{P1} & \textbf{P5} & \textbf{P6} & \textbf{P7} & \textbf{P8} & \textbf{P13} & \textbf{P15} & \textbf{P18} & \textbf{P19} & $\boldsymbol{\mu \pm \sigma}$ \\
\midrule
Canvases & 1 & 2 & 2 & 4 & 2 & 1 & 4 & 2 & 1 & 3 & 2.2 ± 1.1 \\
\midrule
\multicolumn{12}{l}{Concepts} \\
\quad Added & 60 & 38 & 32 & 287 & 51 & 33 & 70 & 91 & 33 & 47 & 74.2 ± 77.1 \\
\quad Removed & 16 & 0 & 0 & 22 & 0 & 0 & 48 & 62 & 0 & 2 & 15.0 ± 22.7 \\
\quad Zoomed & 59 & 0 & 0 & 0 & 0 & 0 & 75 & 0 & 0 & 2 & 13.6 ± 28.4 \\
\midrule
\multicolumn{12}{l}{Hypotheses} \\
\quad Generated & 12 & 12 & 24 & 36 & 12 & 12 & 54 & 25 & 12 & 24 & 22.3 ± 13.9 \\
\quad Saved & 13 & 4 & 5 & 35 & 10 & 12 & 20 & 21 & 3 & 16 & 13.9 ± 9.7 \\
\midrule
\multicolumn{12}{l}{Volunteered} \\
\quad Theories & 0 & 0 & 0 & \textbf{2} & 0 & \textbf{1} & \textbf{4} & \textbf{1} & 0 & \textbf{5} & 1.3 ± 1.8 \\
\quad Queries & 0 & \textbf{1} & 0 & \textbf{2} & 0 & 0 & 0 & \textbf{1} & 0 & 0 & 0.4 ± 0.7 \\
\bottomrule
\end{tabular}
\end{table}

Participants brought theory questions from their own research (Table~\ref{tab:study-participants}), most (7/10) from a project already underway: P13 planned to test the hypotheses in fieldwork the following week, and P18, early in a study of AI tutoring, was ``very happy to be proven wrong'' by literature that would ``change our experimental design.'' All reported learning something from the system. They averaged 117.8 active minutes (SD=51.7) across multiple sittings and days (9.5 sessions; SD=3.4), and most (6/10) voluntarily built additional graphs or theories beyond the one required (Table~\ref{tab:study-engagement}).

\begin{figure*}[t]
  \centering
  \includegraphics[width=0.85\linewidth]{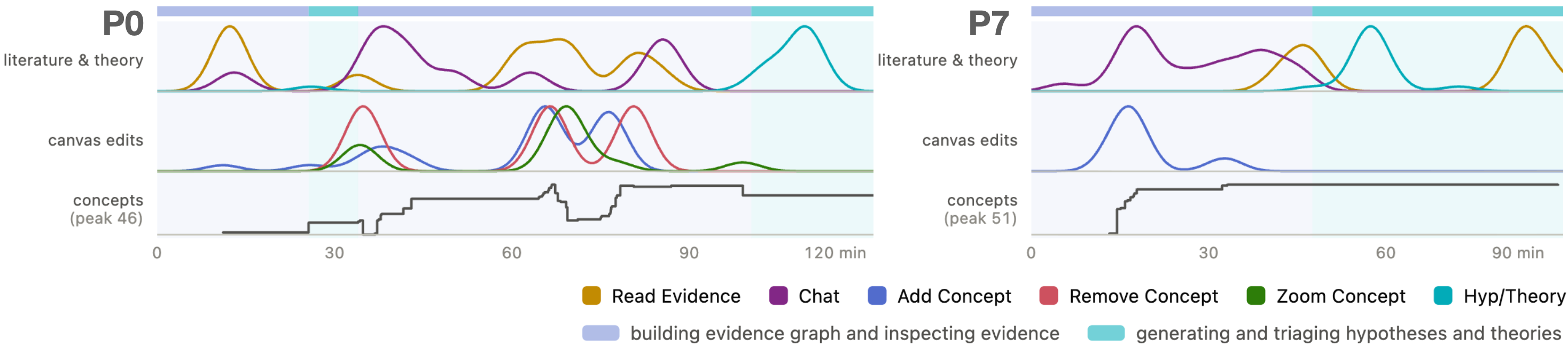}
  \caption{Two common strategies interacting with \sysname{}.  \textbf{Top:} how often they took different kind of action (smoothed over 3 minutes; with each curve scaled to its own peak.) \textbf{Bottom:} the number of concepts on their canvases. The horizontal axis is engaged time, concatenating multiple sessions. P0 added concepts more aggressively, and curate further by removing and zooming them; P7 added concepts more selectively, and never removed any. Both inspected evidence throughout, and P7 additionally went back to evidence on the canvas after theory creation, while P0 mostly read the evidence included in the theory report (not shown).}
  \Description{Interaction traces of P0 and P7 demonstrate two common strategies using \sysname{} each with a horizontal axis of engaged minutes running from zero to about 120. Each panel holds an upper plot of six smooth colored curves, one per action type, and a lower plot with a dark step line for the number of concepts on the canvas. A colored strip along the top of each panel alternates between two shades marking building the evidence graph and generating and triaging hypotheses. In P0's panel the six curves rise and fall repeatedly through the first hundred minutes, and the concepts line climbs to a peak, dips, recovers, and falls again near the end. In P7's panel the curves are sparser, the concepts line rises early and then stays flat, and the amber Read Evidence curve peaks again near minute eighty-five, inside the stretch marked as hypothesis work. A legend below the two panels names the six action colors: Read Evidence, Chat, Add Concept, Remove Concept, Zoom Concept, and Hyp/Theory.}
  \label{fig:interaction-traces}
\end{figure*}

We saw two common strategies participants used when interacting with \sysname{} (Figure~\ref{fig:interaction-traces}). As examples, P0 added widely and then cut back, placing 60 concepts and removing 16 as the canvas rose to 46 and ended at 36, while P7 was selective from the start, placing 51 and removing none. Both read the evidence behind their concepts while building the graph, and both turned to hypotheses in the second half of the session. P7 went back to the graph for evidence afterwards, while P0 read the evidence included in the theory report.

\subsubsection{Participants valued personal relevance and prior knowledge over evidence weight (DG2, DG3)}
\label{sec:findings-curation}

Concepts can be added on the canvas by one of four routes (Table~\ref{tab:study-provenance}). Browsing the ranked list of extracted relations; expanding an existing concept to cover its relational and ontological neighbors; a pre-generated hypothesis brings all the concepts it cites; the chat agent proposes a shortlist of concepts then triples. On average, 66\% of concepts came through the agent's proposals, 21\% through hypotheses, 7\% through neighbors, and 6\% through browsing the full triple list. %

Based on the interviews and interaction logs, participants began by adding widely. Most (7/10) started with chat and accepted most of the agent proposed with minor tweaks. And all used the canvas and chat to learn what the corpus held before building toward a theory (DG1). P0 described this process and diverging then converging: they ``[I] first wanted to just diverge and add as many nodes as possible on the canvas.'' P6 explained why she added so widely, since she could ``be good at one domain, but I mean, not an expert on everything. So I'm pretty open to add new relations.'' The logs agree, since removals came mostly from four participants clearing a canvas to start over, while the other six removed at most two concepts (Table~\ref{tab:study-engagement}). Most (6/10) said the canvas taught them new relations between concepts they already knew. P1, for example, had not known that ``\cpt{knowledge distillation} was connected to \cpt{mode collapse} and also \cpt{reward hacking}.'' Some (6/10) also named something the corpus lacked, from a missing context to missing their own papers, often due to lack of open-access licensing.

Participants decided what stayed on canvas primarily based on their relevance to their project and their prior knowledge, and rarely based on how many papers were behind a concept. Most (9/10) sought concepts they knew mattered or judged the agents proposals against familiarity of the field, and most (6/10) pruned by relevance: ``if they are something that not related into my thesis that I'm working on, I usually just delete them'' (P13). While they do read the paper counts on each node and edge (DG4) but seldom influenced their decisions to keep or remove; only P13 reported removing a concept because it was ``supported by not so much paper, maybe only two paper.''

The canvas sometimes grew faster than participants could read it. Most (7/10) felt overwhelmed at some point, and four (P7, P13, P15, P18) asked the agent to clean up, group, or delete for them, which it cannot, since it only proposes additions (\S\ref{sec:impl-agent}). We had assumed participants would want to make each removal themselves, but they wanted help with removing concepts as much as with adding them. In contrast, the two participants that heavily utilized the zoom feature (\S\ref{sec:zooming-nesting}), P13 and P0 and kept their canvases organized in a few large nested groups (Table~\ref{tab:study-engagement}).

\begin{table}[t]
  \caption{Participants used different methods to add concepts, ranging from \emph{browsing} the full list of triples to find familiar concepts, to anchoring on a node to see its relational and ontological \emph{neighbors}, to adding the concepts mentioned in a pre-generated \emph{hypothesis}, and to asking the \emph{chat} agent to propose concepts or triples.}%
  \label{tab:study-provenance}
  \scriptsize
  \setlength{\tabcolsep}{3pt}
  \centering
  \begin{tabular}{lrrrrrrrrrrr}
\toprule
 Source & \textbf{P0} & \textbf{P1} & \textbf{P5} & \textbf{P6} & \textbf{P7} & \textbf{P8} & \textbf{P13} & \textbf{P15} & \textbf{P18} & \textbf{P19} & \textbf{\%} \\
\midrule
Browse List & 14 & 0 & 1 & 6 & 0 & 2 & 10 & 0 & 0 & 6 & 6 ± 8 \\
Concept-Neighbors & 10 & 0 & 0 & 9 & 1 & 0 & 28 & 1 & 1 & 0 & 7 ± 13 \\
Concept-Hypotheses & 9 & 0 & 0 & 192 & 0 & 0 & 0 & 35 & 22 & 13 & 21 ± 27 \\
Chat & 27 & 38 & 31 & 80 & 50 & 31 & 32 & 55 & 10 & 28 & 66 ± 29 \\
\quad /Proposed & /50 & /80 & /52 & /124 & /61 & /69 & /74 & /99 & /27 & /34 & \\
\bottomrule
\end{tabular}
\end{table}

\subsubsection{Participants chose their level of abstraction, on the canvas and in the hypotheses (DG2)}
\label{sec:findings-abstraction}

Participants were generally satisfied with the abstraction level the system had chosen, and reasoned about how general or specific a concept or hypothesis should be. The system nests proposed concepts by the ontology, so most canvases held nested groups. For the participants who did utilize the zoom feature, P13 wanted their many specific technologies to read as one but worried that ``if I only put tourism technology, I think I lost all the nodes in here, [so I nested them instead].'' P0 treated zooming out as ``an AI-assisted affinity diagram''. Zooming out of \cpt{AI Agent Embodiment} offered its parent \cpt{AI Agent Design} (12 concepts) and the top-level \cpt{Design Factor} (72); P0 took the middle, since ``design factor is like, then too broad. Like it doesn't really make sense.''

When the ontology lacked the grouping a participant wanted, they grouped by locations on the canvas. P7 arranged the canvas by relation structure instead of by ontological structure, with upstream interventions, mechanisms, and downstream outcomes in three columns, and P15 asked the agent to separate positive from negative framings ``so I can better develop the hypothesis later.'' P1 tried to go one level finer than the canvas allows (DG2), asking the agent whether a mixed increase-decrease relation ``would it be different in different domains''; the canvas cannot split a relation by the context of its studies, but the agent's summary ``was able to distinguish different type of domains.'' On the other hand, hypothesis generation was able to this level of granularity with hypotheses generated over disagreeing evidence described paradoxes and inverted-U relations, and most participants (5/10) kept, exported, or named one at least one hypotheses that focused on contradicting or context-specific relations as a main takeaway.

Participants often judged hypotheses by how general and specific they were (9/10), which P0 named: ``there's like a sweet spot for the level of abstraction.'' A hypothesis ``very specific to a device that was introduced in the study'' was dismissed by P15, who kept an ambitious one because ``nobody has sort of explicitly made the contribution this way [...] the way they framed it is less ambitious.'' The logs show the same pattern, as hypotheses built mostly on the ontology's most specific concepts were passed over more often (LeafConcepts, below).

\begin{table}[t]
  \caption{The model's outcome and four hypothesis properties.}
  \label{tab:triage-terms}
  \footnotesize
  \setlength{\tabcolsep}{3pt}
  \centering
  \begin{tabular}{@{}lp{0.68\columnwidth}@{}}
    \toprule
    Term & Definition \\
    \midrule
    Kept (outcome)  & Hypothesis marked as interesting or added to a theory\\
    Breadth & \# of canvas relations cited\\
    LeafConcepts & \% of cited concepts at the ontology's most specific level\\
    AnchorPapers & \# of papers behind the best-supported relation cited\\
    AnchorAgreement & \% of its relations in the majority direction\\
    \bottomrule
  \end{tabular}
\end{table}

\begin{figure*}[t]
\centering
\begin{minipage}{.49\textwidth}
  \centering

  \includegraphics[width=0.7\linewidth]{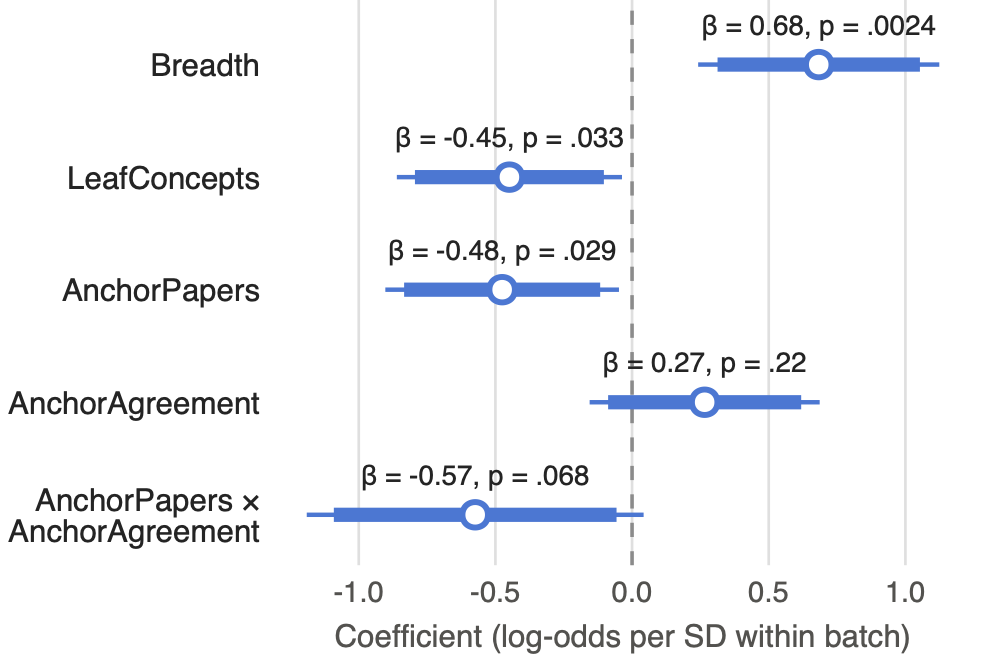}
  \caption{Coefficients of four properties of generated hypotheses and one interaction, from a mixed-effects logistic regression over the 143 hypotheses in batches where a participant both kept and passed over some (thick bars 90\% CI, thin bars 95\% CI). A positive coefficient means the property made a hypothesis more likely to be kept than the others in its batch, holding the rest fixed.}
  \Description{A dot-and-whisker plot with five rows, Breadth, LeafConcepts, AnchorPapers, AnchorAgreement, and the AnchorPapers by AnchorAgreement interaction. Each row shows a point estimate with a thick 90 percent and a thin 95 percent confidence interval on a log-odds axis from minus one to one, annotated with its coefficient and p value. Breadth sits to the right of zero at 0.68 with p equal to .002; LeafConcepts at minus 0.45 with p equal to .033 and AnchorPapers at minus 0.48 with p equal to .029 sit to the left; AnchorAgreement at 0.27 with p equal to .22 straddles zero; the interaction at minus 0.57 with p equal to .068 lies mostly to the left.}
  \label{fig:triage-coefficients}

\end{minipage}
\hfill
\begin{minipage}{.49\textwidth}
  \centering

  \includegraphics[width=0.7\linewidth]{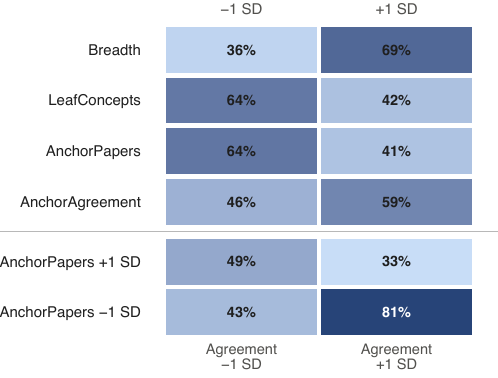}
  \caption{Probability that a generated hypothesis is kept, from the same model, for a typical participant and batch; with every property at its batch mean, it is 53\%. Top four rows: each property one standard deviation below and above its batch mean, with the others at the batch mean. Bottom two rows: a heavy ($+1$ SD) or light ($-1$ SD) best-supported relation crossed with divided ($-1$ SD) or agreed ($+1$ SD) evidence behind it, AnchorAgreement taking the two columns.}
  \Description{A heat-map grid of percentages with two columns, minus one standard deviation and plus one standard deviation, and six rows in two groups. The first four rows, Breadth, LeafConcepts, AnchorPapers, and AnchorAgreement, read 36 and 69 percent; 64 and 42; 64 and 41; and 46 and 59. Below a horizontal rule, two rows labelled AnchorPapers plus one standard deviation and AnchorPapers minus one standard deviation, whose columns are AnchorAgreement at minus and plus one standard deviation, read 49 and 33 percent for the heavy relation, 43 and 81 percent for the light one.}
  \label{fig:triage-probabilities}

\end{minipage}
\end{figure*}

\subsubsection{Participants kept hypotheses that were surprising and covered more concepts and relations on their evidence graph (DG3, DG4)}
\label{sec:findings-triage}

To see what separated the hypotheses participants kept from those they dismissed, we conducted a regression analysis with factors based on what we had learn from the interviews. The ten participants generated 223 hypotheses in 19 batches and reviewed 184. 100 was marked as interesting and 84 passed over or dismissed. We fit a mixed-effects logistic regression predicting whether a hypothesis was kept from four properties, with participant and generation batch as random effects (Table~\ref{tab:triage-terms}):\footnote{R package \texttt{lme4}~\cite{88513115}}
{\scriptsize
\begin{align*}
\text{Kept} \sim{}& \text{Breadth} + \text{LeafConcepts} + \text{AnchorAgreement}\\
& + \text{AnchorPapers} + \text{AnchorPapers} \times \text{AnchorAgreement}\\
& + (1 \mid \text{Participant}) + (1 \mid \text{Batch})
\end{align*}}%
Figure~\ref{fig:triage-coefficients} reports the coefficients and Figure~\ref{fig:triage-probabilities} the probability of keeping a hypothesis as each property varies.

In the interviews, the most common reason for keeping or dismissing a hypothesis was personal relevance (9/10); P5: ``I was trying to find a small subset on which I can kind of tell a story for my paper.'' The next most common reason was novelty. Most (7/10) kept hypotheses that were surprising and new to them, and some (6/10) dismissed hypotheses as already known. P18 and P19 both dismissed a hypothesis matching their own unpublished results, ``it's my ongoing project so I actually did some experiment on that and this is not something surprising''. P6 kept one that introduced a term she had not met before, which she valued because it would ``expand my horizon a little bit, but still connect with what I currently know.'' Results from the regression analysis had similar results where the more papers stood behind the best-supported relation a hypothesis cited, the less often it was kept (AnchorPapers: $\beta{=}{-}0.48$, $p{=}0.029$).

Hovering over a hypothesis highlights the part of the evidence graph behind it (DG4), and participants also favored hypotheses that drew on more of their canvas. P13 used that footprint as a criterion, dropping hypotheses ``only affecting a really small proportion of the graph.'' Hypotheses citing more of the canvas's relations were kept more often (Breadth: $\beta{=}0.68$, $p{=}0.002$).

Finally, hypotheses built mostly on the ontology's most specific concepts were kept less often (LeafConcepts: $\beta{=}{-}0.45$, $p{=}0.033$), matching the too-specific hypotheses participants dismissed in the previous subsection.

Participants also paid close attention to conflicting evidence. \sysname{} ranks relations with many agreeing sources first, and participants relied on those to understand their corpus. However, once they were hypothesizing, most (7/10) attended to evidence that pulled in opposite directions. For example, P8 ``not only included the 100\% agree part'' but also ``the disagreement part because we want to see both sides,'' so as not to ``bias it in one direction'' and to see ``how would my theory or hypothesis come up with.'' For P0, a generated paradox joined two intuitions they had held separately under one mechanism, that ``the same mechanisms that make external member valuable for a group [...] simultaneously create conditions that erodes human engagement'': ``I didn't have the word of social presence [...] So I was never able to derive this part, that it's paradoxical relationship.''

Two participants who doubted a paradox were convinced by the extracted evidence behind it (DG1). P18 kept one hypothesis as ``kind of a contrarian take'' to ``see if this is actually right or is the system [...] just cherry picking stuff,'' exported it, and found it ``reasonably convincing'' once the report cited ``empirical studies actually done on this'' and set the claim's scope.
The regression shows the same attention to conflicting evidence, and where it applied. Whether the papers behind a hypothesis's best-supported relation agreed did not matter on its own (AnchorAgreement: $\beta{=}0.27$, $p{=}0.22$), but its interaction with AnchorPapers improved the fit ($\chi^2_1{=}4.1$, $p{=}0.042$; AnchorPapers $\times$ AnchorAgreement: $\beta{=}{-}0.57$, $p{=}0.068$). Moving AnchorPapers from $-1$ to $+1$ SD lowered a hypothesis's probability of being kept from 81\% to 33\% when the papers behind that relation agreed, and left it close to unchanged (43\% to 49\%) when they did not (Figure~\ref{fig:triage-probabilities}). \emph{In sum, participants prioritized hypotheses that covered many concepts on their canvas, cites fewer leaf concepts, and that a heavily studied relation lowers the chance of a hypothesis being kept when the papers behind it agreed, but not when they were divided.}

\section{Study 2: Case Studies with Domain Scientists}
\label{sec:case-studies}

In contrast to the field deployment, whose participants were incentivized, early-career computer science researchers, our case studies focused on understanding how \sysname{} performs in  with domain senior researchers working in different fields.
We followed the principle of Multi-dimensional In-depth Long-term Case Studies (MILC) proposed by Shneiderman and Plaisant~\cite{207160702}.
The two case studies ran for 2--4 months, with two groups of agricultural and immunology researchers. We initially reached out offering them a new AI tool that explores literature to build hypotheses and theories. The groups were not paid and were free to stop at any point.

Each engagement started with us gathering queries from the researchers and operating \sysname{} ourselves to produce theories reports to share with them. The two teams eventually accessed the system themselves, and the agriculture researchers went on to build graphs and theories of their own from corpora they assembled. We met the agriculture group three times and the immunology group five times over video call for 30--60 minutes. In the meetings they shared their screens and walked us through how they used \sysname{}. All meetings were recorded and transcribed with their consent, and the summaries below were reviewed and approved by them.

Several findings from the deployment recurred with the domain scientists. For example, curating the evidence graph based on personal preferences and project context, including known concepts and discovering new ones. when asked about how they triaged the hypotheses, A1 said it was mostly based on personal preferences, and I2 read a heavy evidence count as a reason to pass over a hypothesis because ``If there's too much evidence for it, somebody's probably shown it and we're going to get feedback [from reviewers] that says, well, we like already know this.'', the pattern AnchorPapers picked up in the regression analysis.
The rest of this section focuses on the distinct learning from the case studies.

\subsection{Agriculture Researchers}
\label{sec:case-agriculture}

\sysname{} was first demonstrated to six researchers at a global agricultural research partnership with a research presence in more than 20 countries across Africa, Asia, and Latin America.
Two of the six researchers chose to bring us their own questions: A1, a bioinformatics scientist and statistician who works on coffee, and A2, an agronomist who leads a team building a conservation indicator for potato landraces in the Andes (Table~\ref{tab:study-participants}).

\subsubsection{Direct comparison to Theorizer on the same query.}
We had A1 make a direct comparison between \sysname{}'s and Theorizer's~\cite{284981954} output on the same theory query. 
He preferred \sysname{}, calling it ``much clearer and more insightful,'' and put the difference down to seeing the evidence as concept-relations, which ``helped me follow the reasoning [behind a hypothesis] without having to take out a pencil and paper to connect all the pieces of information myself.'' When describing a hypothesis he found useful, he mentioned concepts and ontological hierarchies: ``It was quite interesting to see that this \emph{kind of phenomenon} occurring [in] a \emph{group of plants}, but there are also \emph{another group of plants} that \emph{this [a relationship] doesn't hold}, and the report [theory] mentioned it.''

\subsubsection{A1 checked the theory against the literature and against his own knowledge.}
Participants in the case studies were using \sysname{} for their active research projects, and put in more effort to verify its outputs than the deployment participants, who mostly checked hypotheses against the evidence \sysname{} itself showed them (\S\ref{sec:findings-triage}). A1 verified it against public protein databases and reported that ``everything that I read on the report is correct.''
He returned most often to the three fields that state a hypothesis's limits, its domain scope, its novelty evaluation, and its special cases.
On literature discovery and understanding, he found the report also useful for writing the discussion section of a paper, ``because you cannot read 200 papers in one week. It's impossible.''

\subsubsection{A1 valued parts of theories that pointed at experiments.}

A1 was drawn to the \emph{unaccounted for} section of the theory (\S\ref{sec:impl-theory}), which listed seven mechanisms in the evidence graph that the theory did not explain, among them \cpt{ABA signaling} and \cpt{unfolded protein response}. He knew both concepts well, and both link to \cpt{heat-shock response}; neither, however, had been on his mind while reading. ``Because you see it, you think of it, and you have more ideas and more ways to discuss your findings.'' Reading a later report, on how \cpt{C.~arabica} arose from its two parent species, he identified a specific experiment he would propose, a growth-chamber temperature gradient measuring how \cpt{C.~canephora} and \cpt{C.~arabica} expression ratios shift as temperature rises, and estimated it at around ten thousand US dollars. He also judged that one or two such experiments, run against the predictions of both reports together, could support roughly six new papers.

\subsubsection{Interacting with \sysname{}.}
Using \sysname{} on their own, A1 asked how telomere-to-telomere sequencing is helping resolve the genomes of polyploid plants,
and A2 continued to focus on landraces, and queried ``build a theory of how to create a Quantitative Red List of crop landraces.''%

The most consistent theme was that both found things in their own graphs that they had not known. A1 pointed at a node for \emph{hifiasm}, a genome assembler: ``I never heard of them before constructing the graph, which is great.'' The theory he saved a week before our meeting was organized around it, opening on a hifiasm-centered assembly paradigm and citing evidence from his canvas for each clause. %

Similarly, A2's graph contradicted a model his team is actively building. Their indicator scores how well a landrace is conserved in a community by combining four standard measures: how many families hold the variety (OCF), how much of it each of them grows (RCF), and how far it is spread geographically and across altitude. One hypothesis \sysname{} generated said the four do not compensate for one another (\S\ref{app:hyp-stewardship}). When OCF falls to a single custodian household, it argued, no achievable level of the other three can compensate, because losing that one maintainer removes the landrace from cultivation entirely. The theory A2 then synthesized stated the general form, that the four dimensions ``interact in fundamentally non-additive, threshold-dependent, and asymmetrically compensatory ways,'' and named the additive scoring built into existing threat assessment frameworks as the assumption it was contradicting. A2, who leads that work, had not considered it: ``if it's just one family at the end, it does not matter if the geographic dispersion or the altitude in the dispersion or the RCF, the cultivation abundances, for that variety, because one family has it. [...] I didn't think about this.'' %

\subsubsection{Reading the canvas as ideas instead of papers.}
Once the chat agent had grown A1's canvas well past what he was willing to review by hand, he read it structurally. Noticing that two competing long-read sequencing platforms both connected to the same downstream assembly-quality nodes, he drew a practical conclusion about how to spend a sequencing budget: ``now that I see multi platforms, it means, when you do your analysis or your project, you won't say, I want to spend only on this company. I want multiple approaches because you will have the results.'' The hypothesis open beside him made the same argument in prose and listed eight observations behind it, among them one reporting that neither platform alone produced a gap-free kiwifruit assembly, and it proposed combining both platforms.
When asked whether curating the canvas was about understanding the papers or about arriving at a hypothesis, A1 said it was closer to \emph{ideation}: ``I see this as ideas rather than [...] making connections. I see this as ideas. And then I collect all the ideas that I find, I found interesting.''

\subsubsection{Where \sysname{} fell short.}
The clearest gap was literature coverage in the agriculture domains (\S\ref{sec:findings-curation}). Beyond a missing concept (\emph{Coffea eugenioides}, a parent species of \emph{C.~arabica} central to his second question, was absent from the graph), much of what A1's field knows about coffee varieties is recorded in curated genome databases and not in the prose of papers, and \sysname{} reads only papers. For example, he pointed to one aggregating sequence data across roughly 50 coffee varieties, and noted that while its records ultimately trace back to publications, the curated form is what researchers actually consult and it is not reachable through the paper search our pipeline builds on. Extending extraction to structured community databases, or linking concepts in the graph to their records, is the change he asked for most directly.

\subsection{Immunology Researchers}
\label{sec:case-immunology}

A second case study was conducted with a team of immunology researchers from a large nonprofit research institute. Their queries carry more structure than a free-text question. Each names a small set of differentially expressed genes and a downstream phenotype, and the graph fills in the mechanisms between them.
For this case study we built a specialized extraction pipeline tuned with biological ontologies (e.g., \cite{282589077}). The specialized backend is outside this paper's scope, but the team used the same interface described in \S\ref{sec:system}.

This early engagement ran over two months. Two of the researchers explored the system on their own as well as during our meetings, thinking aloud as they worked. The goal of this engagement was to explore early promise and pitfalls in this direction and to lay out a roadmap for continued collaboration.
I2, the immunologist who leads the group, named the team's bottleneck: ``...our hardest part right now is which hypotheses should we follow up with wet bench [experiments]? Because that's a huge investment of our time and resources.''

\subsubsection{Interactive canvas versus static theory reports.}
We sent this team PDF reports before giving them access to the system. The first report on I2's cell type predicted close to the opposite of what her team expected, and she attributed it to her own query instead of the pipeline: ``we only gave you 10 genes versus the hundreds of genes that we use in our pathway analysis.'' In chat she could supply what the graph had never been given, and the answers moved: ``because I put in extra like pieces of information that I know, but it wouldn't know because it only has the 10 genes [...] it seems to be giving us outcomes that are more aligned than that PDF was.'' She also ran hypothesis generation over her canvas and read the six hypotheses it returned, though she did not triage them or synthesize a theory before the meeting ended (DG3).

\subsubsection{\sysname{} proposed mechanisms outside their expertise and outside their standard analysis.}
The step I2's team takes after differential expression is hallmark pathway analysis, which scores their genes against a curated collection of 50 pathways. \sysname{} placed \cpt{TLR4 signaling} on her canvas, a concept that collection does not contain, and she recognized it immediately: ``I really like this one popping up [...] because we can stimulate monocytes with [TLR4], but [TLR4] wasn't coming up in our hallmark analysis.'' I1, a computational biologist on the team, pointed out the reason, that hallmark is ``only 50 gene sets,'' and I2 agreed the graph was ``actually more encompassing.'' This is the gap we measured against survey outlines in \S\ref{sec:tech-eval}, now in a wet lab. A curated collection covers what a field has agreed on, and the concepts a particular inquiry needs can fall outside it (DG1). While deployment participants mostly learned new relations between concepts they already knew (\S\ref{sec:findings-curation}), here the graph surfaced a mechanism the team's standard analysis could not have. I1 valued the same in a theory \sysname{} generated on drug efflux, ``probably not a direction that I would have thought to go on my own, maybe because I just don't know a ton about [...] drug efflux.'' Asked by I2 whether the two transporters that theory named could be treated as one family, he zoomed both to their shared ontology parent and kept them nested inside it, pooling their evidence under \cpt{ABC transporters} while the individual transporters stayed on the canvas (DG2; \S\ref{sec:findings-abstraction}).

The team read more of the evidence behind a hypothesis when the experiment to test it would be expensive. I2 skipped it for TLR4, where she already knew the biology and the test was cheap: ``this is useful if true [...] we can actually say, oh, great, we can test this.'' %

\subsubsection{Experiments, novelty, and feasibility.}
\sysname{} reports the number of unique sources behind a relation as a measure of how well supported it is. Choosing where to spend months of bench time, I2 read that number in the opposite direction: ``we want to prioritize things that haven't been shown before, but are very likely to be true [...] if there's too much evidence for it, somebody's probably shown it and we're going to get feedback that says, well, we like already know this.'' This is the reading the regression in \S\ref{sec:findings-triage} found among the deployment participants.
Some of their criteria for triaging hypotheses were specific to their lab, such as whether the target sits on the cell surface and whether reagents can be bought. I1 also wanted the confidence shown on a relation to carry biological provenance and not only a count, since a relation observed in T cells generally may not hold in the subset they study. He then argued against his own request, since narrowing to a cell type would ``miss those things that could be happening that haven't been observed.''
Feasibility and resources also shaped which hypotheses were worth an experiment, a criterion absent from the deployment, where participants judged hypotheses by relevance, novelty, and breadth (\S\ref{sec:findings-triage}). Reading a hypothesis that predicts children would show a larger fold change in IL-6 than adults, I2 said ``[this] actually is the type of readout we would want [...] maybe we'll just have to do this experiment,'' then described the experiment: take frozen cells, isolate the monocytes, stimulate them with LPS in a culture dish, and assay the supernatant for IL-6. She concluded by judging the experiment to be inexpensive and feasible. %

\section{Discussion}

\subsection{Iterative Design Retrospective}

Design and development of \sysname{} spanned eleven months, including early pilot testing and co-design sessions. Several early pilot users became coauthors, forming an interdisciplinary team of HCI, NLP, and biology researchers and designers. Below we document the qualitative learning from this iterative design process, including changes we made from observing particular failure modes in previous versions to inform the designs of future theory development support systems.

\subsubsection{Balancing users between exploring broadly vs deeply was an early design tension.} 
In the first few months, \sysname{} extracted fewer concepts from fewer papers and displayed all nodes and edges by default, expecting this to provide a broad overview first and let users trim away less interesting ones. The design quickly became overwhelming as the system scaled. We therefore moved to starting with a blank canvas and designed several different ways for users to explore and \emph{discover} concepts.

Several interaction patterns were tested, and some made it to the final design. Early attempts included allowing users to explore and directly add concepts from the ontology (Figure~\ref{fig:ontology}), or explore a paper list and add all triples extracted from individual papers. This, however, led to users overly focused on understanding individual papers and the cliques around them. The ranked list of every extracted relation (\S\ref{sec:all-triples}) was useful when users wanted to filter by keyword and quickly add known concepts. However, the sheer number of relations in the list does not support discovery well. We therefore allowed users to anchor on a concept on canvas and explore relevant concepts in ways described in the methods section. To support discovering concept further away from an anchor, we added pre-generated hypotheses which allowed users to discover additional concepts that fits with the anchor to form hypotheses (\S\ref{sec:expanding-from-node}).
Finally, chat was the most flexible when users had vague high-level or nuanced intent, and we eventually combined it with in-chat GUI widgets to scaffold concept discovery.%

\subsubsection{Users were more selective of concepts than of relations.} Early pilots suggested users were fairly exhaustive in including edges between nodes they deemed important.
The edges served more as affordances to summarize and access the evidence between concepts. At times, the canvas could become cluttered with an many edges. We updated the visualization so that edges are de-emphasized and progressively de-saturated by zoom level when not selected. \sysname{} also auto-accepts edges between any two added concepts if their unique-source count clears a user-adjustable threshold. These changes helped, but participants in our field deployment still reported being overwhelmed by relations (\S\ref{sec:field-study}), and we suspect the visual de-emphasis should be more aggressive. At the same time, many participants in the field study still reported discovering unexpected and useful relationships between known concepts.%

\subsection{Study Limitations}

\subsubsection{User study.} Our field deployment asked participants to work on their own questions at their own pace, spending multiple hours with the system over one to two weeks (\S\ref{sec:field-study}). An alternative would be a controlled lab study with two shorter sessions comparing \sysname{} to a baseline. We deliberately opted for depth over control. Theory building is cognitively demanding, and we wanted participants to spend multiple hours deeply engaging with the literature, hypotheses, and theories, and allow them to do so in multiple sessions at their own pace to avoid fatigue. A lab study with multiple conditions would lead to them only spending a fraction of that time. %
We pushed further toward realism with two unincentivized case studies, with groups of agricultural and immunology researchers using the system on their own research questions (\S\ref{sec:case-studies}).

\subsubsection{Theory quality evaluation.}
Our hypothesis and theory generation components are adapted from Theorizer~\cite{284981954} (\S\ref{sec:impl-hypothesis}, \S\ref{sec:impl-theory}), which was validated in a large-scale knowledge-cutoff evaluation showing that language models can generate theories predictive of results reported in later-published papers.
We changed how the evidence is compiled from papers, from faceted evidence extraction (detailed in \cite{284981954}) to aggregated concept-relation edges carrying each paper's claim, observation, and experimental design (\S\ref{sec:impl-hypothesis}). Rerunning the same benchmarks would further validate our approach but is costly and beyond the scope of this study. More fundamentally, our evidence structure is designed to support interactivity, not fully autonomous theory generation. We instead checked the quality of the extracted evidence against 12 expert-curated survey outlines in a technical evaluation (\S\ref{sec:tech-eval}), and the full interactive system in our deployment and case studies with researchers from different domains (\S\ref{sec:field-study}, \S\ref{sec:case-studies}).

\subsection{System Limitations}

\subsubsection{Static base graph.}
The base graph from the extraction pipeline is static during user interaction, but users may wish to \emph{change} or \emph{expand} it. For example, a researcher might want to search and import additional evidence on a particular relation, either between existing or new concepts. Or, they might wish to group concepts differently by inserting new custom concepts into the ontology.
Allowing users to enrich the base graph during synthesis also raises new research questions: \emph{How should a system help researchers fluidly switch between gathering evidence and synthesizing it while theorizing?}

\subsubsection{Visualization: Depth and Context}
While most participants found the graph visualization useful, researchers in domains where the concept-relation structure is shallow find it less useful.
Specifically, domain focusing on methods to hill-climb on shared benchmarks. In such cases, the graphs end up shallow with many techniques fanning into few metrics. For example, P8 saw little value in the graph visualization and asked why the graph was needed, and agreed when we suggested that a list of concepts might have served them better, describing their work as ``more breadth than the depth part.''

Our current visualization also do show the context a finding was held under. For example, P1 asked the agent whether a relation whose papers disagreed ``would it be different in different domains'' (\S\ref{sec:findings-abstraction}). To illustrate with an example, a paper that reported 
\triple{Approach A}{\edgeup{outperforms}}{Approach B} may have additional context such as ``\emph{by S on benchmark T when U}'' or detailed experiment setup in a few sentences. In our current implementation, the canvas cannot split a relation by the context. Such context is kept in the free form evidence snippet and experiment design fields. The chat agent and hypotheses generator can synthesize over them. For example, P1 was able to get a summary of the contexts behind a diverging edge, and, eventually a hypothesis about a potential paradox behind them. Further visualizing such context is challenging, due to how sparse and difficult to normalize they can be.

\subsection{Future Work and Opportunities}

\subsubsection{Domain-specific extraction and synthesis.}
The case studies made clear that domain specialization opens new research opportunities. Our pipeline infers its ontology from the papers and from the model's parametric knowledge, a reasonable default for fields without a curated shared vocabulary, such as those in our deployment.
However, many communities in the natural sciences maintain such vocabularies, and our inferred ontology misses concepts they expect when the open-access literature is thin (\S\ref{sec:case-agriculture}). Equipping a domain-specialized pipeline with community-curated ontologies and commonly used relations is ongoing work. Similarly, \sysname{} measures evidence strength by unique sources count, yet different domains have their own ways of measuring and reporting the uncertainty of experimental results; in biology, this could mean the specific experimental methodology.
More generally, future work could extract the detailed statistical analyses behind each experiment and conduct meta-analyses across the many experiments behind each edge.

\subsubsection{Further developing a single hypothesis.}
\label{sec:limit-depth}
After finding a hypothesis valuable, some participants wished to further develop it, especially when considering devoting resources to run an experiment. A1 said they wanted a long report per hypothesis. They imagined the system doing an exhaustive literature search around the hypothesis, and to see it sharpen or fail based on new evidence. Currently, users could create a new, more focused graph to iterate on a single hypothesis but lose the bigger context. Future work could explore designs where users could \emph{zoom} into one hypothesis and zoom back out to a broader graph.

\subsubsection{From hypotheses to experiments to new evidence.}
\label{sec:discussion-experiments}
An exciting direction for future work is to extend theorizing into designing and running experiments that collect new observations, completing the loop. Our case studies showed early promise for this direction: A1 read theories as pointers to experiments and as a filter over data he had already collected (\S\ref{sec:case-agriculture}), and I2, reading a hypothesis about IL-6, immediately sketched the inexpensive experiment that would test it (\S\ref{sec:case-immunology}). Expanding a hypothesis into concrete experimental designs, and ranking those designs by how feasible they are given a researcher's resources, would let a researcher theorize around what they can actually test; this matters most in domains where experiments are prohibitively costly, which is why feasibility surfaced as a triage criterion in the case studies and not in the deployment (\S\ref{sec:case-immunology}).
For computational fields, a natural extension is to combine \sysname{} with data-driven discovery systems, handing a hypothesis off as the specification of a computational experiment.

Conversely, since our participants were using \sysname{} to support their active research project, they often mentioned new findings from their current experiments and studies. While they primarily use it to triage the concepts and hypotheses in the system, future work could allow users to incorporate such pre-published experimental results into the system as new concept-relation triples and included as evidence for generating the next rounds of hypotheses.

\section{Conclusion}

In this paper, we presented \sysname{}, which extracts the thousands of experimental observation from up to hundreds of papers as concept-relation triples, normalized and organized by generating an ontology. This gives researchers control over theory building at scale by exploring the literature and curating an evidence graph of concepts and relations they find interesting, and then triaging hypotheses candidates to form a provisional theory.

We evaluated it in a study with computer science researchers and in two case studies with immunology and agriculture researchers. Participants curated their evidence graphs by relevance to their own projects and preferences, more than by how well supported a relation was. They chose the level of abstraction they worked at, on the canvas and in the hypotheses, and saved hypotheses that surprised them, covered more of their graph, and made sense of evidence that disagreed. The domain scientists found mechanisms outside their standard analyses and hypotheses they judged potentially worth an experiment.

Theorizing today is so costly that few researchers attempt it at scale. Our work expands prior work in ideation support systems that typically only considers grounding a specific idea in a few papers, and builds on prior work in computational scientific discovery that are fully autonomous. 
Doing so fills a long-standing gap in the literature between automated scientific discovery and a recent call to further interactive theorizing tools that allow researchers to draw on their intuition, preferences and domain knowledge to guide theory synthesis. \cite{15548809,278062769}.

We imagine a future where theory synthesis in hours is an \emph{ideation probe at scale}, part of everyday research instead of a grand status to be scrutinized and defended. A researcher who finds the evidence thinner than expected can discard the theory and try many other framings, fluidly aggregate evidence around different concepts at different levels of abstraction against the vast landscape of literature. 
The hard part is then no longer synthesis but judgment, deciding which concepts to build on and which theory is worth an experiment.

\bibliographystyle{ACM-Reference-Format}
\bibliography{references}

\clearpage
\appendix
\section{Example Generated Hypotheses}
\label{app:example-hypothesis}

Two hypotheses generated by the system, reproduced as the interface presented them. The first is discussed in \S\ref{sec:intro-example}, the second in \S\ref{sec:case-agriculture}.

\subsection{Asymmetric Error Correction Advantage}
\label{app:hyp-asymmetric}

\begin{quote}
\emph{Example-based explanations produce a net positive effect on team performance despite increasing both overreliance and underreliance because the magnitude of performance gains from improved AI error detection exceeds the magnitude of performance losses from increased rejection of correct AI predictions. Quantitatively, the error detection improvement (e.g., +12-20 percentage points in correct rejection of AI errors) is typically larger than the accuracy loss on correct AI predictions (e.g., -4 to -5 percentage points from underreliance), yielding net complementary team performance.}
\end{quote}

\paragraph{Domain scope}

\begin{quote}
\emph{Human-AI teams performing classification tasks where AI accuracy is substantially below 100\% (typically 50-80\%), such that there are sufficient AI errors for humans to detect, and where example-based explanations (nearest neighbors, prototypes) are provided alongside AI predictions.}
\end{quote}

\paragraph{Special cases}

\begin{quote}
\emph{When AI accuracy approaches 100\%, the asymmetric advantage diminishes because there are few errors to detect, and underreliance costs dominate.}

\emph{In fine-grained adversarial settings where examples are indistinguishable to lay users, the error detection advantage disappears.}

\emph{The advantage is strongest when example quality/diversity is high enough to reveal AI errors but not so high as to cause excessive general skepticism.}
\end{quote}

\paragraph{Supporting evidence}

\begin{quote}
\emph{On CUB, CHM-Corr users correctly reject 53.45\% of misclassified images vs.~41.22\% for no-explanation users (+12.23pp gain on errors) \cite{251224004}, while losing only \textasciitilde4pp on correct predictions \cite{251224004}, producing net positive team performance.}

\emph{Example-based explanations achieved complementary performance in income prediction (71.1\% vs.~61.1\% human-alone vs.~62.5\% AI) and biography classification (71.2\% vs.~60.0\% vs.~62.5\%) \cite{255998499}, while feature-based explanations did not improve over no-AI baseline.}

\emph{3-NN was the only explanation type to achieve complementary performance on fine-grained dog classification (82.88\% vs.~81.14\% AI-only) \cite{235253892}, demonstrating that example-based explanations uniquely enable net gains.}

\emph{On adversarial ImageNet images where AI is always wrong, 3-NN achieved 75.07\% error detection accuracy with a consistent +4\% gain over confidence-only \cite{235253892}, confirming superior error detection capability.}

\emph{Example-based explanations showed no decrease in participant accuracy when AI was wrong ($\beta$=$-$0.03, p=0.80 for income; $\beta$=0.01, p=0.43 for biography), while feature-based caused significant decreases \cite{255998499}.}
\end{quote}

\subsection{Stewardship Singularity Threshold}
\label{app:hyp-stewardship}

Generated over A2's landrace graph (\S\ref{sec:case-agriculture}). OCF, RCF, GDF, and ADF are the labels the generated text uses for the four measures his team's conservation indicator combines.

\begin{quote}
\emph{When household stewardship redundancy (OCF) falls to a single custodian household or family within a locality, the landrace enters a critical extinction-risk zone where no achievable level of geographic dispersion (GDF), altitudinal breadth (ADF), or cultivation abundance (RCF) can compensate for the stewardship deficit, because the loss of that single maintainer---through aging, migration, livelihood change, or disaster---eliminates the landrace from active cultivation irreversibly. Below approximately 7 maintaining households per community (a participatory threshold identified in four-square analyses), landrace persistence becomes increasingly fragile and stochastic rather than deterministic.}
\end{quote}

\paragraph{Domain scope}

\begin{quote}
\emph{Applies to on-farm conservation of crop landraces maintained through farmer seed-saving in smallholder farming systems, particularly in regions where landraces are not backed up in functional ex situ collections. Most directly evidenced in South and Southeast Asian contexts (India, Nepal) but generalizable to any system where landraces depend on active human cultivation rather than wild persistence. Does not apply to landraces already secured in accessible, viable genebank collections from which reintroduction is feasible.}
\end{quote}

\paragraph{Special cases}

\begin{quote}
\emph{Culturally or religiously linked landraces may persist at very low OCF due to non-economic motivations, but even these can cross the extinction threshold when cultural practices erode.}

\emph{Clonally propagated crops (starchy roots/tubers) face even sharper thresholds because seed cannot be easily stored or exchanged informally.}

\emph{Post-disaster contexts can abruptly push OCF below threshold across multiple communities simultaneously.}
\end{quote}

\paragraph{Supporting evidence}

\begin{quote}
\emph{On-farm conservation sites and custodian farmers were rare and spatially isolated; even farmers immediately adjacent to conservation farmsteads were not cultivating landraces, meaning discontinuation by a single custodian family could lead to irreversible loss \cite{269716933}.}

\emph{Six landraces in Western Odisha were restricted to single villages, maintained by few households on small patches, classified as rare or isolated \cite{285893149}.}

\emph{Several Eastern Odisha landraces including Pokia Dhan, Pandkagura, Para Dhan were found in very limited geographies with low conservation scores, often restricted to single villages \cite{285893149}.}

\emph{The four-square participatory analysis used a threshold of 7 out of 15 households to distinguish `many' from `few' maintainers, classifying landraces below this as vulnerable \cite{285893149}.}

\emph{Unique landraces with distinctive traits such as perennial growth habit can be completely lost without timely conservation intervention, as demonstrated by Garve Cauli's extinction \cite{228960238}.}

\emph{Farmer migration causes landrace loss by reducing the number of active maintainers below viable levels \cite{267826293, 228977719}.}

\emph{The landrace threat assessment methodology explicitly incorporates maintainer number reduction as a key threat criterion \cite{267826293}.}
\end{quote}

\section{Chat Agent Tools}
\label{app:tools}

Table~\ref{tab:agent-tools} lists every tool \sysname{}'s chat agent can call, in the order the interface documents them to users: first the tools that find concepts and relations in the extracted graph, then the ones that propose changes to the canvas, then the ones that read the evidence behind a relation, and finally the ones that track the state of the session and the theory being built. The same list is available to users in the interface, through a capabilities panel that shows each tool alongside the verbatim system prompt and the canvas snapshot the agent receives each turn.

The \emph{Runs} column records where each tool executes. Only \texttt{ask\_paper} runs on the server, because it queries a full-text service the browser cannot reach. The remaining ten are declared to the model server-side but dispatched in the browser against live interface state, and their results are returned to the server-side loop; the ranked triple pool, the canvas, and the candidate hypothesis list are all client state, so running them anywhere else would mean keeping a second copy in sync. No tool writes to the canvas: the ones that would instead render a card the researcher accepts or dismisses.

\begin{table*}[t]
  \caption{The eleven tools available to \sysname{}'s chat agent. Relation labels in the ranked triple pool are derived from the extracted evidence, so a tool may choose which concept pairs to propose but never what relation holds between them.}
  \label{tab:agent-tools}
  \footnotesize
  \setlength{\tabcolsep}{8pt}
  \renewcommand{\arraystretch}{1.1}
  \begin{tabularx}{\textwidth}{@{}lX@{}}
    \toprule
    \textbf{Tool} &  \textbf{Purpose} \\
    \midrule
    \multicolumn{2}{@{}l@{}}{\textbf{\textsc{Finding concepts and relations}}} \\
    \addlinespace[0.3em]
    \texttt{search\_suggested\_triples} & Keyword-search the ranked pool of triples that can be added to the canvas, matching against source and target labels, their aliases, and the relation text. Returns up to 50 triples ranked by how many papers back the relation and how far they agree on its label; called with no arguments it returns the whole addable pool. Triples conflicting with concepts already on the canvas are filtered out. \\
    \addlinespace[0.4em]
    \texttt{expand\_around\_node} & Walk one hop outward from a single concept, returning the top-ranked pool triples whose other endpoint is not yet on the canvas. Restricted to incoming relations, outgoing relations, or both. \\
    \addlinespace[0.4em]
    \texttt{find\_paths\_between} & Find directed chains of pool triples connecting two concepts through intermediates, for questions of the form ``how does A lead to B'' and for bridging disconnected regions of the canvas. \\
    \addlinespace[1em]
    \multicolumn{2}{@{}l@{}}{\textbf{\textsc{Proposing changes to the canvas}}} \\
    \addlinespace[0.3em]
    \texttt{propose\_concepts} & Render a card of up to eight concepts, one clickable row each with a one-line gloss from the agent; the label and paper count on each row are computed from the data rather than written by the model. Clicking a row adds that single concept and opens a relations widget around it. \\
    \addlinespace[0.4em]
    \texttt{propose\_add\_triples} & Stage three to eight triples as a proposal card with a one-sentence framing and a per-triple rationale, and a button stating the change to the canvas. The agent chooses the concept pairs; each pair's relation is looked up in the pool. Two triples sharing a chain identifier are proposed as one pathway the researcher accepts as a unit. \\
    \addlinespace[1em]
    \multicolumn{2}{@{}l@{}}{\textbf{\textsc{Reading the evidence}}} \\
    \addlinespace[0.3em]
    \texttt{get\_finding\_details} & Read the claim, observation, and experiment design behind one extracted finding cited in the canvas snapshot. \\
    \addlinespace[0.4em]
    \texttt{ask\_paper} & Ask a targeted question against a paper's full text and return the answer with supporting excerpts and page numbers. Used when the extracted findings do not cover what was asked. \\
    \addlinespace[1em]
    \multicolumn{2}{@{}l@{}}{\textbf{\textsc{Session and theory state}}} \\
    \addlinespace[0.3em]
    \texttt{set\_goal} & Record a short statement of what the researcher is building toward, re-injected into the canvas snapshot on every subsequent turn. \\
    \addlinespace[0.4em]
    \texttt{suggest\_next\_messages} & Offer one to three follow-up messages as clickable pills below a response, each sent verbatim as if the researcher had typed it. \\
    \addlinespace[0.4em]
    \texttt{get\_user\_theory} & Report where the researcher is in building a theory: no theory yet, triaging candidate hypotheses (returning each hypothesis with its scope and its keep-or-dismiss label), or reading a finished theory. \\
    \addlinespace[0.4em]
    \texttt{filter\_theory} & Render a widget that narrows the candidate hypothesis list to a named theme, one call per theme. Valid only while the researcher is triaging hypotheses. \\
    \bottomrule
  \end{tabularx}
\end{table*}

\section{Triple Ranking and the Default Zoom Level}
\label{app:triple-ranking}

\sysname{} maintains a single ranked pool of candidate relations, used by the list of extracted relations described in \S\ref{sec:all-triples}, by the triage widget the chat agent returns (Figure~\ref{fig:triples}), and by the agent's own search tools (Table~\ref{tab:agent-tools}). Below we describe how this pool is built.

We first lift each extracted triple up the ontology: a triple between two concepts is also a candidate relation between any pair of their ontology ancestors, including the two concepts themselves, and we merge the evidence of every triple that falls under the same pair. We then drop the pairs in which one concept is an ancestor of the other, and those supported by fewer than two unique papers.
We also exclude top-level ontology concepts (Figure~\ref{fig:ontology}). %

We then score each remaining pair of concepts $(S, T)$ as follows:
\[
\beta \ln(1 + p) + \gamma c - \lambda\,[\,\ln(1 + i_S) + \ln(1 + i_T)\,] - \mu\,(h_S + h_T)
\]

\noindent In the equation, $p$ is the number of unique papers behind the merged evidence, $c$ is the proportion of those papers reporting the pair's most common relation, $i_S$ and $i_T$ are the numbers of extracted relations that connect two descendants of $S$ and of $T$, which using $S$ or $T$ as an endpoint would hide, and $h_S$ and $h_T$ are the heights of the two concepts, measured as the longest path from each down to a leaf of the ontology. The first two terms prefer relations that are supported by more papers and that those papers agree on. The last two push each relation down toward the most specific concepts that still carry its evidence. We used $\beta{=}0.6$, $\gamma{=}0.8$, $\lambda{=}0.5$, and $\mu{=}0.2$ as the default weights, and \sysname{} also exposes them and the source threshold in the interface for users to adjust.

Finally, the same underlying relation can score well at several levels of abstraction at the same time. We consider two candidates to be duplicates when their source concepts are in an ancestor-descendant relationship and their target concepts are as well. Going down the ranked list, we drop a candidate when a higher scoring duplicate has already been kept. Each relation is therefore listed only once, at the pair of concepts where it scored best, and this is also the level of abstraction \sysname{} first presents it at on the canvas (\S\ref{sec:all-triples}).

\section{Evidence Sampling for Hypothesis Pre-generation}
\label{app:evidence-sampling}

Phase 5 (\S\ref{sec:impl-hypothesis}) generates hypotheses over the whole graph before anyone opens it. However, a graph built from a hundred papers serializes to more evidence than fits in one prompt. This appendix describes the evidence sampling strategy over multiple generation rounds.%

We first format every aggregated edge in the graph as an evidence item carrying its observations, experimental setups, and source papers.
Within a round, each edge $e$ in the evidence pool carries a weight combining how well attested it is with how heavily it has already been used:

\[ w_e \;=\; s_e^{0.8} \;\cdot\; \left(1 + c_e\right)^{-1.2} \]

\noindent where $s_e$ is the number of unique papers supporting the edge and $c_e$ the number of hypotheses generated so far that cite it. The first factor prefers well-evidenced edges. %
The second factor keeps successive runs from converging on the same evidence. It decays without reaching zero, so a heavily cited edge always retains some chance of being drawn again. The decay is also steep enough for uncited edges to overtake cited ones. Without this property the sampler would keep re-showing its best-supported edges.

Each run is given a token budget for evidence

\[ B \;=\; \min\left(865{,}000,\; 0.35 \times \lvert \text{evidence\_pool} \rvert \right) \]

\noindent The first term is a technical ceiling: a one-million-token context less the 32,000 tokens reserved for the completion, roughly 3,000 for the system prompt and template, and a 100,000-token safety margin. The second term caps any single prompt at just over a third of the graph, so no run can see enough of it to duplicate another.

One generation is three sequential rounds of two parallel runs, each run producing eight hypotheses, for 48 in total.
Sampling decides only which edges a prompt contains, and within the prompt they are ordered by source count in descending order, so that the strongest evidence comes first. %

\section{Sanity Checks under Zhu et al.'s Protocol}
\label{app:sanity-check}

Zhu et al.~\cite{281496866} score a single \emph{partition} of the papers, where each paper is assigned to its single most-immediate concept and the resulting partition is compared to the survey's using three standard cluster-agreement metrics (NMI, ARI, and Purity). Our ontology is not a partition, since each paper attaches to many concepts. \S\ref{sec:tech-eval} therefore scores survey concepts one at a time instead. The method we ran is their best-performing one, which prompts an LLM to propose candidate aspects at each level of the hierarchy and clusters paper summaries under them. We ran it and our own pipeline on the same models (\texttt{Claude Opus 4.6} and \texttt{text-embedding-3-large}), in place of the GPT-4o and LLaMA-3.1-8B split their paper uses. We clustered with $k$-means, which their released code uses although the paper describes a Gaussian mixture, and a Gaussian-mixture ablation scored lower on all three metrics. The protocol supports two sanity checks on the comparison. First, our run scores within $\pm$0.04 of the published TaxoBench-CS results on our 12-ontology subset (NMI 0.622, ARI 0.185, Purity 0.585), which suggests the model substitution is a small effect. Second, if a perfectly guided user picked the closest-matching concept from our ontology for each survey concept, the resulting partitions would score on par with theirs on all three metrics (Table~\ref{tab:oracle-partition}).

\begin{table}[t]
  \caption{Partition-level scores under Zhu et al.'s protocol (higher is closer to the survey ontology), with 95\% CIs over the 12 ontologies.}
  \label{tab:oracle-partition}
  \footnotesize
  \centering
  \begin{tabular}{lrrc}
    \toprule
    Metric & Ours & Zhu et al. & Per-ontology higher\\
    \midrule
    NMI    & 0.666 [0.597, 0.735] & 0.622 [0.557, 0.687] & 8 / 12 \\
    ARI    & 0.255 [0.183, 0.327] & 0.185 [0.101, 0.270] & 10 / 12 \\
    Purity & 0.644 [0.560, 0.728] & 0.585 [0.506, 0.665] & 6 / 12 \\
    \bottomrule
  \end{tabular}
\end{table}

\section{Hypothesis Triage Model: Fit and Checks}
\label{app:triage-model}

The full model fit better than one with only the random effects (likelihood ratio $\chi^2_5{=}17.9$, $p{=}0.003$). Two further fixed effects, a hypothesis's rank in the review list and the share of its relations backed by a single paper, explained no further variance ($\chi^2_1{=}0.71$, $p{=}0.40$; $\chi^2_1{=}0.74$, $p{=}0.39$). Every coefficient keeps its sign under a conditional logistic regression stratified by batch (interaction $\beta{=}{-}0.64$, $p{=}0.081$), under leave-one-participant-out refits, and under a strict coding that counts only explicit dismissals as negatives, in which LeafConcepts is the only term that stays significant; quadratic and spline terms for AnchorPapers are null. Finally, AnchorAgreement is shrunk toward the corpus base rate in proportion to how few findings produced it, an empirical-Bayes estimate with a $\mathrm{Beta}(7.69, 2.16)$ prior fitted by maximum likelihood to the 387 cited relations with two or more directional findings. As a result, a 50\% agreement among 12 sources counts as a bigger split than a 50\% agreement among 2.

\section{Prompts}
\label{app:prompts}

For reproducibility, the rest of the appendix lists all prompts \sysname{} used in Jinja2 format.%

\subsection{Finding Extraction}
\label{app:prompt-findings}

One prompt serves both extraction passes (\S\ref{sec:impl-extraction} and \S\ref{sec:impl-linking}). On the first pass the ontology block is empty; on the second the constructed ontology is rendered into it and the model is additionally asked to link every triple endpoint to an ontology concept.

\begin{lstlisting}[style=prompt]
You are an expert scientific research assistant and a librarian specializing in systematic literature review and knowledge graph construction.

Your task is to extract the following information given a research paper:
- A list of the specific experiments conducted in the paper, including the details of the experiment designs.
- For each experiment, a list of detailed and specific observations that were made in the experiment.
- For each observation, a corresponding claim that was made by the authors based on the observation.
- For each of the above, also provide the corresponding evidence snippets from the paper so other researchers can verify the information you extracted without reading the entire paper.


<IMPORTANT GUIDELINE: PIPELINE CONTEXT>
Your extracted triplets will be used downstream to build a causal knowledge graph across many papers. For the biology domain, this will be nodes and edges that forms larger bio pathways.

In this graph:
- Each triplet's `source_concept` and `target_concept` will become **nodes** in the causal graph.
- The `relation` will become a directed **edge** between them.

Because of this, source and target concepts must be meaningful, specific, nameable concepts — the kind of thing you would label a node in a graph. Good concepts include: model names (e.g., "GPT-4"), method names (e.g., "Chain-of-Thought Prompting"), metrics (e.g., "BLEU score"), phenomena (e.g., "hallucination"), tasks (e.g., "question answering"), datasets (e.g., "MMLU"), or other well-defined entities.

**Do NOT** use long descriptions or sentence fragments as source/target concepts. If you find yourself writing a multi-clause phrase as a concept, you likely need to split it into multiple triplets with simpler concepts.

The `context` field is for experimental conditions or setup details that qualify the relationship (e.g., "on the SQuAD dataset", "with 5-shot prompting") — NOT for hiding important concepts that should be in source/target.
</IMPORTANT GUIDELINE: PIPELINE CONTEXT>

Your response should be a JSON object with the following structure:
{
  "experiments": [
    {
      "name": "The name of a specific experiment from the paper. (e.g. 'Experiment 1', 'Offline Evaluation', etc.)",
      "design": "A description of the design of the experiment. Include details such as number of participants, recruitment methods, dataset size, etc.",
      "evidence": ["Exact quote 1 from the paper", "Exact quote 2 from the paper"]
      "observations": [
        {
          "evidence": ["Exact quote 1 from the paper", "Exact quote 2 from the paper"]
          "observation": "A detailed and specific observation that was made in the experiment",
          "triplets": [ # based on the observation, evidence, and the full paper, create triplets of source concept, target concept, and relation. See below for detailed guidelines.
            {
              "source_concept": "The concept that is the source of the claim or observation",
              "target_concept": "The concept that is the target of the claim or observation",
              "relation": "The relation between the source and target concepts",
{% if ontology_json %}
              "source_ontology_node_id": "ont_1", # the ontology node_id matching the source_concept, or null if no match
              "target_ontology_node_id": "ont_5", # the ontology node_id matching the target_concept, or null if no match
{% endif %}
              "context": "A brief description of the context, either from the method of the paper or the experimental design, that cannot be easily captured by the source and target concepts, and is specific to the current claim or observation."
            }
          ],
          "claim": "A corresponding claim that was made by the authors based on the observation",
          "gap": "A brief description of how strongly the observation supports the claim. Start with '[DIRECTLY SUPPORTS]', '[INDIRECTLY SUPPORTS]', '[SPECULATES]', '[OPINIONS]' and then describe the reason for this relation in 1-2 sentences."
        }
      ]
    }
  ]
}

Guidelines:
- Extract between 3-15 observations depending on the paper's content
- Each observation should be 1-3 sentences summarizing a key result or conclusion
- Prioritize observations from the Results, Discussion, and Conclusion sections
- Include quantitative results when available (e.g., "achieved 95% accuracy")
- Include 1-3 evidence excerpts per observation and per experiment
- Evidence should be *EXACT* text spans copied from the paper

<IMPORTANT GUIDELINE: OBSERVATIONS AND CLAIMS>

- Observations are concrete things that had happened during an experiment and are specific to the experiment.
- When extracting such observations, you should include specific details about the experiment setup and what lead to the observation. For example, instead of saying "combining 4 methods leads to ..." you should include details about what the 4 methods are and how they were combined in the experiment. This way, someone could read what you have written and reproduce the observation without reading the entire paper.
- Claims, instead, are phrased by the authors in a more general way than the specific observations. They are often phrased as generalizable findings that are useful for informing future research ideas.
{% if ontology_json %}
- In REFLECTION MODE, claims should still be rephrased to be understandable without reading the paper. However, you should use the ontology concept labels for specific names when they appear in the ontology (e.g., use "GPT-4" from the ontology rather than rephrasing it away). The ontology already provides the normalized vocabulary that connects these names across papers.
{% else %}
- You should also rephrase the claims when needed so that they can be understood by other researchers without reading the entire paper. Claims should generalize the observation into broader insights. NEVER say "current paper", "current studies", or "the proposed method".
- However, triplets from claims should still use specific, nameable concepts as source/target (as described in the PIPELINE CONTEXT section above). If the claim generalizes a finding about "GPT-4" to "large language models", the claim triplet should use "Large language models" as the concept — not a long descriptive phrase.
{% endif %}
- IMPORTANT: Make sure to be comprehensive and include EVERY observation and claim that are made in the paper for each experiment. Do not miss any.
- When appropriate, include the paper's weaker results and limitations to include as observations and claims. This should not include general limitations of the paper, such as dataset size, participant demographics, etc., but rather specific observations and claims the authors made as limitations or future work.
- Capture ABSOLUTELY EVERY OBSERVATION made during each experiment that are related to any claims made by the authors.

</IMPORTANT GUIDELINE: OBSERVATIONS AND CLAIMS>

<IMPORTANT GUIDELINE: TRIPLETS>
- Triplets capture the causal relationships identified in the observations and in the claims.
- Generate at least 1 triplet for each observation and each claim.
- The source and target concepts in the triplets should be specific. For example, if the paper tested GPT-3.5, use "GPT-3.5" as the concept name, and NOT "LLM".
- If the paper tested multiple models, e.g., GPT-3.5 and GPT-4, create one triplet per model and do not try to combine them into a single one.
- For example, if the observation is that "X increases Y", the triplet should be:
  - Source concept: "X"
  - Target concept: "Y"
  - Relation: "increases"
- If the observation is that "X decreases Y", the triplet should be:
  - Source concept: "X"
  - Target concept: "Y"
  - Relation: "decreases"
- If the observation is that "X is related to Y", the triplet should be:
  - Source concept: "X"
  - Target concept: "Y"
  - Relation: "is related to"
- Observations and claims can be summarized at different levels of granularity. You should also include triplets for the different levels of granularity.
- IMPORTANT: Always use the most specific concept name available as source/target. If a method, model, or dataset has a name, use that name (e.g., "GPT-4", "LoRA", "SQuAD"). Do NOT create separate generic-description triplets restating the same concept without its name — generalization is handled downstream by the ontology hierarchy.
{% if ontology_json %}
- In REFLECTION MODE: The ontology already provides normalized names for specific concepts. Use the ontology labels directly in your triplets (see the ONTOLOGY REFLECTION section below).
{% endif %}
- When an observation mentions multiple specific entities (e.g., methods, models, datasets), generate a separate triplet for EACH entity. For example:

Observation: "[Method M] with [Mechanism Y] performed the highest [Metric Z] in [Context C], outperforming [Mechanism S] by 10% on [Dataset T] and by 12% on [Dataset U]."

Should generate four triplets (source, relation, target, context):
(M, increases, Z, C with T and U)
(Y, increases, Z, C with T and U)
(M, outperforms, S, C with T and U)
(Y, outperforms, S, C with T and U)

- Example 1:
  - Observation: "Including specific descriptions about the goals and intents of a user will help LLMs better align with the user's goals and intents, but decreases its creativity and ability to help users discover new ideas."
  - Triplet 1:
    - Source concept: "Goal-specific prompting"
    - Target concept: "LLM creativity"
    - Relation: "decreases"
    - Context: "When user goals and intents are described in detail"
  - Triplet 2:
    - Source concept: "Goal-specific prompting"
    - Target concept: "Goal alignment"
    - Relation: "increases"
    - Context: "In LLM outputs"
  - Claim: "Well specified task description will help AI models better align with pre-compiled gold standard answers, but constrains its ability to generate novel solutions, especially in open-ended tasks."
  - Triplet 1:
    - Source concept: "Task specification detail"
    - Target concept: "Gold standard alignment"
    - Relation: "increases"
    - Context: "In AI model outputs"
  - Triplet 2:
    - Source concept: "Task specification detail"
    - Target concept: "Novel solution generation"
    - Relation: "decreases"
    - Context: "Open-ended tasks"
- Example 2: (observational)
  - Observation: "Users switch between browser tabs to maximize their learning efficiency, especially when they have exhausted most of the information in the current tab."
  - Triplet:
    - Source concept: "Tab switching"
    - Target concept: "Learning efficiency"
    - Relation: "increases"
    - Context: "When users have exhausted most information in the current tab"
  - Claim: "Designs that allow users to efficiently manage and switch between information sources will help maximize their learning rate."
  - Triplet:
    - Source concept: "Information source management"
    - Target concept: "Learning rate"
    - Relation: "increases"
    - Context: None
- IMPORTANT: **Never use collective or vague group references as concepts.** If the paper says "all models", "SOTA models", "all methods", "existing approaches", or similar, you MUST replace these with the actual specific entities the authors tested or observed. The authors are referring to a specific set — look at the experiment design and results to identify exactly which ones. Generate a separate triplet for each.

  Bad example (DO NOT do this):
  - Observation: "All models showed improved accuracy on the benchmark after fine-tuning."
  - Triplet: ("All models", "increases", "Accuracy") ← WRONG: "All models" is vague

  Good example:
  - Observation: "All models showed improved accuracy on the benchmark after fine-tuning."
  - (The experiment tested BERT, RoBERTa, and DeBERTa)
  - Triplet 1: ("BERT", "increases", "Accuracy", "After fine-tuning on benchmark")
  - Triplet 2: ("RoBERTa", "increases", "Accuracy", "After fine-tuning on benchmark")
  - Triplet 3: ("DeBERTa", "increases", "Accuracy", "After fine-tuning on benchmark")

  Another bad example:
  - Observation: "SOTA models struggle with multi-hop reasoning tasks."
  - Triplet: ("SOTA models", "decreases", "Multi-hop reasoning performance") ← WRONG

  Good example:
  - (The paper evaluated GPT-4, Claude-3, and Gemini Pro on multi-hop reasoning)
  - Triplet 1: ("GPT-4", "decreases", "Multi-hop reasoning performance")
  - Triplet 2: ("Claude-3", "decreases", "Multi-hop reasoning performance")
  - Triplet 3: ("Gemini Pro", "decreases", "Multi-hop reasoning performance")

- IMPORTANT: Notice that the relations should be short phrases and does not include detailed context about the specific observations; the additional context should be captured by the concepts and contexts fields.
- IMPORTANT: If the observation or claim statements have conjunctions (e.g., "and", "or", "but"), you should generate a triplet for each conjunction.
- IMPORTANT: !!! Exhaustively list all of the causal relations in both the observation and claim.

</IMPORTANT GUIDELINE: TRIPLETS>

{% if ontology_json %}
<IMPORTANT GUIDELINE: ONTOLOGY REFLECTION>

You are running in REFLECTION MODE. This paper was previously analyzed along with other papers in this research area, and the ontology below was generated from the combined findings. You are now re-extracting observations with the benefit of this shared vocabulary so that the output integrates well with the ontology and produces higher quality data for a causal graph.

## Ontology-Guided Extraction Rules

1. **Base triplets ONLY on observations**: Every triplet must be grounded in an actual observation from the paper. Do NOT invent triplets just to match ontology concepts. The observation and evidence must come from the paper text.

2. **Reuse ontology concepts explicitly**: When an observation involves a concept that exists in the ontology, you MUST use the exact same terminology as the ontology label for the source_concept or target_concept in your triplets. For example:
   - If the ontology has a node labeled "GPT-4" and the paper discusses GPT-4, use "GPT-4" (not "GPT4" or "gpt-4" or "OpenAI's model").
   - If the ontology has "Chain-of-Thought Prompting" and the paper discusses CoT, use "Chain-of-Thought Prompting" as the concept name.

3. **Generate separate triplets for each specific concept**: If the observation mentions multiple specific entities that each exist in the ontology, generate a separate triplet for EACH one. For example:
   - If the observation says "GPT-4 and Claude-3.5 both outperform smaller models on reasoning tasks" and the ontology has nodes for "GPT-4", "Claude-3.5", and "Reasoning Tasks", generate separate triplets for GPT-4 and for Claude-3.5, rather than combining them.
   - If the observation mentions 3 specific datasets from the ontology, generate triplets for each dataset separately.

4. **Maintain specificity**: Use the most specific ontology concept available. If the ontology has both "Large Language Model" (intermediate) and "GPT-4" (specific), and the paper specifically discusses GPT-4, use "GPT-4" in your triplets.

5. **Link triplet source/target to ontology nodes**: For EACH triplet, set `source_ontology_node_id` and `target_ontology_node_id` to the `node_id` of the best-matching ontology concept for that source/target concept. This creates a direct link from each part of the triplet to the ontology.
   - If the source concept matches an ontology node exactly (e.g., source_concept is "GPT-4" and ontology has a node labeled "GPT-4"), use that node's `node_id`.
   - If the source concept is a more specific variant not in the ontology, link to the closest parent concept (e.g., if "GPT-4-turbo" isn't in the ontology but "GPT-4" is, link to "GPT-4").
   - If no ontology concept matches the source or target, set the field to `null`.
   - This linking is critical: it enables the downstream causal graph to directly map triplet concepts to the ontology hierarchy.

## Ontology Reference
Below is the hierarchical ontology of concepts. Use the `label` field for concept names in your triplets, and the `node_id` field for the `source_ontology_node_id` and `target_ontology_node_id` fields.

```json
{{ ontology_json }}
```

</IMPORTANT GUIDELINE: ONTOLOGY REFLECTION>
{% endif %}

<Start of paper>

Title: {{ title }}

{{ full_text }}

<End of paper>
<References metadata>

Extract the detailed experiments and their observations from this paper.
\end{lstlisting}

\subsection{Ontology Construction}
\label{app:prompt-ontology}

Run once per corpus over the observations and claims extracted by the first pass (\S\ref{sec:impl-ontology}).

\begin{lstlisting}[style=prompt]
You are ScientistGPT, the most advanced AI scientist in the world. You can answer any scientific question, and if you don't know the answer, you can use your enormous intellect to find it. You answer every question accurately, faithfully, and with the highest level of scientific integrity.

You have an interdisciplinary background. First, you have a deep technical skill in linguistics, specifically for building rigorous ontologies for different domains. You will be reviewing a series of research papers in a different domain that you are also expert in, and you will help build up a new ontology for this field.

For the biology domain, this will be nodes and attributes helps summarizes a large bio pathway graph. For biology data, use the heirarchy relations to describe attributes. For example, gene types, effect types, effected regions, ..., etc.


# Task
Your task is to construct a hierarchical ontology of concepts based on a collection of scientific observations. This ontology will organize and normalize the concepts found across multiple observations into a structured hierarchy. The hierarchy will later be used to normalize nodes before building a causal graph. You will use a combination of the paper data you are reviewing and your domain expertise to figure out the right concepts and hierarchy structure.

## What is a Hierarchical Ontology?
A hierarchical ontology is a structured classification of concepts where:
- **Nodes** represent concepts, variables, or phenomena at different levels of abstraction
- **Hierarchy** represents taxonomic relationships:
  - **"is_a"** (or equivalently **"type_of"**): Classification (e.g., "GPT-4" is_a "large language model").
  - **"part_of"**: Composition (e.g., "attention mechanism" part_of "transformer architecture")

The hierarchy helps organize concepts from specific (e.g., "BERT fine-tuning on SQuAD") to abstract (e.g., "transfer learning").

**IMPORTANT**: This ontology captures ONLY the hierarchical taxonomy of concepts (is-a and part-of relationships). It does NOT capture causal relationships like "increases", "decreases", "enables", etc. Those will be handled separately by the causal graph component.

**IMPORTANT**: The ontology supports nodes with multiple parent nodes. For example, "Python" is-a "interpreted language" and also is-a "oop language". Make sure to leverage this to capture all the concepts in the ontology (both extracted from the papers and also common concepts based on your expert understanding of the domain.)

## How This Ontology Will Be Used: Interactive Causal Graph Viewer

After the ontology is built, a separate step will construct a **causal graph** where nodes represent concepts and edges represent causal relationships (e.g., "model size" → increases → "reasoning ability"). Each causal node will be linked to an ontology node via its `ontology_node_id`.

The ontology powers an **interactive zoom feature** in a web-based causal graph viewer. Users can:

- **Zoom out** on any causal node to collapse it and its siblings into their shared ontology parent. For example, if "GPT-4", "Claude-3", and "LLaMA-2" are all linked to the ontology node "Large Language Model", a user can zoom out to see a single "Large Language Model" node with all their causal edges aggregated.
- **Zoom in** on an aggregated node to expand it back into its children, revealing more specific concepts and their individual causal relationships.
- **Zoom to an abstraction level** to instantly collapse the entire graph to show only abstract, intermediate, or specific concepts.

**Why this matters for your ontology design:**

1. **Intermediate nodes define zoom stops.** Each intermediate node in your hierarchy becomes a level the user can zoom to. If you connect "GPT-4" directly to "AI Model" with no intermediate nodes, the user has no useful zoom level between the specific model and the most abstract category. Adding "Thinking", "Close models" and "GPT Family" as intermediates gives the user three meaningful zoom stops to explore their causal relations to other concepts.


2. **Sibling groupings determine what collapses together.** When a user zooms out to "Large Language Model", ALL causal nodes whose ontology nodes are descendants of "Large Language Model" collapse into one aggregated node. This means your choice of what concepts share a parent directly controls what gets grouped. Good groupings (models that are meaningfully comparable) produce useful aggregations. Bad groupings (unrelated concepts under the same parent) produce confusing aggregations.

3. **Edge aggregation reveals cross-paper patterns.** When nodes collapse, their causal edges merge. If 5 different specific LLMs each have an edge saying "increases → Reasoning Ability", after zooming out they become a single aggregated edge "Large Language Model → increases → Reasoning Ability" supported by evidence from all 5 papers. This is how users discover high-level patterns. Your hierarchy determines which specific findings aggregate into which high-level patterns.

4. **Multiple parents enable multiple zoom paths.** Because a node can have multiple parents, a concept like "GPT-4" can be zoomed out via "Large Language Model" OR via "Reasoning Model" — giving users different analytical perspectives on the same concept.

**Concrete example of the zoom interaction:**

Starting causal graph (specific level):
```
GPT-4 ──increases──→ Code Quality
Claude-3 ──increases──→ Code Quality
LLaMA-2 ──no effect──→ Code Quality
BERT ──decreases──→ Latency
RoBERTa ──decreases──→ Latency
```

With this ontology hierarchy:
```
Language Model (intermediate)
├── Large Language Model (intermediate)
│   ├── GPT-4 (specific)
│   ├── Claude-3 (specific)
│   └── LLaMA-2 (specific)
└── Encoder Model (intermediate)
    ├── BERT (specific)
    └── RoBERTa (specific)
```

After user zooms out to "Large Language Model":
```
Large Language Model ──mixed effects──→ Code Quality  (3 sources: GPT-4, Claude-3, LLaMA-2)
BERT ──decreases──→ Latency
RoBERTa ──decreases──→ Latency
```

After user zooms out further to "Language Model":
```
Language Model ──mixed effects──→ Code Quality  (3 sources)
Language Model ──decreases──→ Latency  (2 sources: BERT, RoBERTa)
```

Notice how the intermediate nodes ("Large Language Model", "Encoder Model") create useful zoom stops that let users explore patterns at different granularities. Without them, the user would jump directly from 5 specific nodes to 1 "Language Model" node with no intermediate view.


# Key Requirements

## 1. Concept Extraction and Normalization
Extract concepts from the triplets in the observations (both source_concept and target_concept) and normalize them.

**CRITICAL: EVERY concept that appears as a `source_concept` or `target_concept` in ANY triplet across ALL observations MUST be represented as a node in the ontology.** Do not skip or omit any concept. If a concept appears in a triplet, it must have a corresponding ontology node (either as its own node, or merged into another node via normalization with the merge documented in `normalization_decisions`). After you finish, verify that every unique concept string from the input triplets maps to at least one ontology node.

Normalization rules:
- **Terminology normalization**: Use consistent phrasing (e.g., choose "large language models" OR "LLMs", not both)
- **Abstraction alignment**: Identify when specific concepts are instances of more general concepts
- **Semantic equivalence**: Merge concepts ONLY when they refer to the exact same thing with different syntax

**CRITICAL: Preserve Specific/Low-Level Concepts**
- **Keep specific entities as leaf nodes**: Model names (GPT-4, BERT, LLaMA-2), specific methods (LoRA, QLoRA, DPO), datasets (SQuAD, MMLU), metrics (BLEU, ROUGE-L), etc. should be preserved as individual leaf nodes.
- **Only merge for syntactic variations**: Merge ONLY when concepts are the exact same thing written differently:
  - "GPT-3" and "gpt3" and "GPT3" → "GPT-3" (same model, different formatting)
  - "BERT-base" and "bert base" → "BERT-base" (same model variant)
  - "chain-of-thought" and "Chain of Thought" and "CoT" → "Chain-of-Thought" (same technique)
- **Do NOT merge semantically related but distinct concepts**:
  - "GPT-3" and "GPT-4" are DIFFERENT models → keep as separate leaf nodes
  - "BERT" and "RoBERTa" are DIFFERENT models → keep as separate leaf nodes
  - "LoRA" and "QLoRA" are DIFFERENT methods → keep as separate leaf nodes
  - "few-shot learning" and "zero-shot learning" are DIFFERENT → keep separate

Examples of concepts that SHOULD be merged (syntactic variations only):
- ["gpt-4", "GPT4", "GPT-4", "gpt 4"] → "GPT-4"
- ["chain of thought", "Chain-of-Thought", "CoT prompting"] → "Chain-of-Thought"
- ["Syntax errors", "syntax error", "Syntatic errors"] → "Syntax Errors"

Examples of concepts that should NOT be merged (semantically distinct):
- "GPT-3.5" and "GPT-4" → keep as separate nodes (different models)
- "fine-tuning" and "pre-training" → keep as separate nodes (different training phases)
- "BLEU" and "ROUGE" → keep as separate nodes (different metrics)

## 2. Building the Hierarchy
Organize concepts into a directed acyclic graph (DAG) hierarchy using ONLY these relationship types:
- **is_a** (or equivalently **type_of**): For classification (e.g., "GPT-4" is_a "large language model")
- **part_of**: For composition (e.g., "self-attention" part_of "transformer")

**IMPORTANT**: Nodes can have MULTIPLE parents. This allows concepts to belong to multiple categories simultaneously. For example, "Python" can be both an "Interpreted Language" (is_a) and an "Object-Oriented Language" (is_a).

Structure:
- **Root nodes**: The most abstract concepts (no parents)
- **Intermediate nodes**: Mid-level abstractions that bridge leaf concepts to roots
- **Leaf nodes**: Specific concepts from the observations

**CRITICAL: You MUST create intermediate bridging nodes.** The input triplets contain mostly specific, leaf-level concepts. Your job is to use your domain expertise to invent the intermediate and abstract nodes that organize them into a proper taxonomy. Do NOT connect leaf nodes directly to root nodes — there should almost always be at least one intermediate node in between. If you see leaf concepts like "GPT-4", "Claude-3", and "LLaMA-3" that all share a common category, you must create intermediate grouping nodes (e.g., "Large Language Model", "GPT Family") even if those terms never appear in the input triplets.

**Target depth**: Most branches should have 3-5 levels (root → intermediate(s) → leaf). A branch with only 2 levels (root → leaf) is almost always too flat. Ask yourself: "Is there a natural mid-level grouping between this leaf and its root?" If yes, create it.

**How to discover intermediate concepts**: Pay close attention to the `claim` field in each observation and the `experiment_design` on the parent experiment — not just the triplets. While triplets contain specific, leaf-level concepts (e.g., "Claude-3.7-Sonnet", "GPT-2"), the author's claim often reveals the higher-level categories the authors are actually reasoning about. The experiment design can also reveal important groupings (e.g., "we compare encoder-only vs decoder-only architectures" suggests those are useful intermediate nodes). These generalizations are excellent intermediate nodes.

For example:
- **Triplet**: `Claude-3.7-Sonnet -outperforms-> GPT-2`
  **Claim**: "Thinking models outperform non-thinking models"
  → Create intermediate nodes: "Thinking Model", "Non-Thinking Model" (both is_a "Language Model"). Then `Claude-3.7-Sonnet` is_a "Thinking Model" and `GPT-2` is_a "Non-Thinking Model".

- **Triplet**: `GPT-4 -achieves higher accuracy than-> BERT-base`
  **Claim**: "Larger models outperform smaller models on reasoning tasks"
  → Create intermediate nodes: "Large-Scale Model", "Small-Scale Model". Then `GPT-4` is_a "Large-Scale Model" and `BERT-base` is_a "Small-Scale Model".

- **Triplet**: `LoRA -reduces memory usage compared to-> Full Fine-Tuning`
  **Claim**: "Parameter-efficient methods match full fine-tuning at lower cost"
  → Create intermediate node: "Parameter-Efficient Fine-Tuning" (is_a "Fine-Tuning Method"). Then `LoRA` is_a "Parameter-Efficient Fine-Tuning" and `Full Fine-Tuning` is_a "Fine-Tuning Method".

- **Triplet**: `MMLU -correlates with-> HellaSwag`
  **Claim**: "Knowledge benchmarks and commonsense benchmarks measure complementary abilities"
  → Create intermediate nodes: "Knowledge Benchmark", "Commonsense Benchmark" (both is_a "Benchmark"). Then `MMLU` is_a "Knowledge Benchmark" and `HellaSwag` is_a "Commonsense Benchmark".

In general: if the claim groups specific entities into a category (e.g., "thinking models", "parameter-efficient methods", "knowledge benchmarks"), that category should become an intermediate node in your ontology.

**Valence-aware grouping for outcome-like concepts (insert when appropriate)**

Some concepts have an inherent **valence** in the context of the domain — they are **good** when they go up, **bad** when they go up, or **neutral** / context-dependent. This applies especially to leaf concepts in the `outcome`, `metric`, and `phenomenon` categories. A single intervention often has effects in both directions simultaneously (e.g. a treatment that *increases* a benefit and *decreases* a harm), and both directions are evidence of the intervention's strength. But if those leaves all share the same intermediate or abstract parent, the downstream causal-graph viewer will aggregate them into a single "mixed effect" edge and the pattern disappears.

When you create an intermediate or abstract parent whose direct children would mix **good-valence** and **bad-valence** leaves, insert an extra valence-grouping layer between that parent and its leaves. Use sibling intermediates named after the valence — e.g. "Beneficial X", "Adverse X", and (only if there are such siblings) "Neutral X". Keep the original parent above the valence layer; do NOT replace it. Inherit the original parent's `category` on the new valence nodes, and assign them `abstraction_level: "intermediate"`.

**Apply this guideline only when:**
1. The valence of each leaf is **unambiguous** in the domain — the field uniformly treats one direction as good and the other as bad.
2. The would-be siblings under one parent actually **contain a mix** of valences (good + bad, or good + neutral, or bad + neutral). If every sibling shares the same valence, no extra layer is needed.
3. The parent is an `outcome` / `metric` / `phenomenon`-style category. Do not apply this to `system`, `method`, `intervention`, `dataset`, `task`, etc. — those rarely have inherent valence.

**Skip this guideline when:**
- The concept's valence is genuinely context-dependent (e.g. "length of stay" — sometimes good, sometimes bad). Place such concepts directly under the original parent, or under a single `Neutral X` group if one already exists for other siblings.
- You would have to invent a fake good/bad/neutral label just to satisfy the rule.

**Example — clinical-trial domain (mixed-valence outcomes).**

Without valence grouping (problematic — high-level aggregation looks contradictory):
```
Clinical Outcome (abstract, outcome)
├── Overall Survival (specific)          ← good
├── Quality of Life (specific)           ← good
├── Disease-Free Remission (specific)    ← good
├── Mortality Rate (specific)            ← bad
├── Drug Toxicity (specific)             ← bad
├── Disease Progression (specific)       ← bad
└── Length of Hospital Stay (specific)   ← neutral / context-dependent
```
A new drug that *increases* Overall Survival and *decreases* Mortality Rate is a clear win, but at the abstract "Clinical Outcome" parent the viewer aggregates both edges into one "mixed effects" signal.

With valence grouping (preferred):
```
Clinical Outcome (abstract, outcome)
├── Beneficial Clinical Outcome (intermediate, outcome)
│   ├── Overall Survival (specific)
│   ├── Quality of Life (specific)
│   └── Disease-Free Remission (specific)
├── Adverse Clinical Outcome (intermediate, outcome)
│   ├── Mortality Rate (specific)
│   ├── Drug Toxicity (specific)
│   └── Disease Progression (specific)
└── Neutral Clinical Outcome (intermediate, outcome)
    └── Length of Hospital Stay (specific)
```
Now the same drug aggregates as `increases → Beneficial Clinical Outcome` and `decreases → Adverse Clinical Outcome` — two consistent, positive signals at the intermediate zoom level.

**Naming.** Prefer domain-natural phrasing: "Beneficial Clinical Outcome" / "Adverse Clinical Outcome", "Performance Benefit" / "Performance Cost", "Desirable Side Effect" / "Undesirable Side Effect". Fall back to "Positive [parent]" / "Negative [parent]" / "Neutral [parent]" when no domain-natural term exists.

For example:
```
AI Model (abstract, root)
├── Language Model (intermediate) [is_a]
│   ├── Large Language Model (intermediate) [is_a]
|   │   ├── GPT Family LLM (intermediate) [is_a]
|   │   │   ├── GPT-4 (specific) [is_a] (multi-parent)
|   │   │   └── GPT-3.5 (specific) [is_a]
│   │   ├── Claude Family LLM [is_a]
|   │   │   └── Claude-3.7-Sonnet (specific) [is_a] (multi-parent)
|   │   └─── Reasoning Models (intermediate) [is_a]
|   │       ├── GPT-4 (specific) [is_a] (multi-parent)
|   │       └── Claude-3.7-Sonnet (specific) [is_a] (multi-parent)
│   └── Small Language Model (intermediate) [is_a]
└── Vision Model (intermediate) [is_a]
    └── ViT (specific) [is_a]

Transformer Architecture (abstract, root)
├── Self-Attention (intermediate) [part_of]
│   └── Multi-Head Attention (specific) [part_of]
└── Feed-Forward Layer (intermediate) [part_of]

Programming Paradigm (abstract, root)
├── Object-Oriented Language (intermediate) [is_a]
│   └── Python (specific) [is_a] (multi-parent)
└── Interpreted Language (intermediate) [is_a]
    └── Python (specific) [is_a] (multi-parent)

Note: some of the nodes have multiple parents:
  Claude-3.7-Sonnet (specific) [is_a Claude Family LLM, is_a Reasoning Models]
  GPT-4 (specific) [is_a GPT Family LLM, is_a Reasoning Models]
  Python (specific) [is_a Object-Oriented Language, is_a Interpreted Language]
```

## 3. Abstraction Levels
Assign each node one of:
- `specific`: Concrete, directly from observations (e.g., "BERT", "SQuAD dataset", "85% accuracy")
- `intermediate`: Moderately general (e.g., "transformer model", "reading comprehension task")
- `abstract`: High-level concept (e.g., "language understanding", "model architecture")

## 4. Categories
Assign each node a category. Use consistent categories across the ontology. Common categories include:
- `method`: Techniques, algorithms, approaches
- `metric`: Measurements, scores, evaluation criteria
- `phenomenon`: Observed behaviors, effects, patterns
- `variable`: Factors that can change or be manipulated
- `intervention`: Actions or changes applied in experiments
- `outcome`: Results, effects, consequences
- `system`: Tools, models, platforms, systems
- `property`: Characteristics, attributes, features
- `task`: Problems, challenges, benchmarks
- `dataset`: Data collections, corpora

You may add other categories if needed, but try to use no more than 8-10 distinct categories.

# Input: Experiments and Observations with Triplets
{% if query %}
Here is the general area/query that your ontology should be centered around:
```
{{ query }}
```
{% endif %}

The input is a list of experiments, each containing multiple observations. Each experiment includes:
- `experiment_design`: How the experiment was set up (use this to discover groupings and categories)
- `observations`: A list of observations from this experiment, each with:
  - `observation`: The specific experimental result
  - `claim`: The author's generalization or hypothesis (use this to discover intermediate concepts)
  - `triplets`: Structured source → relation → target relationships

Extract and organize concepts from the source_concept and target_concept fields of these triplets into a hierarchical ontology:
```
{{ findings_json }}
```


# Output Format
You must return your results in JSON format, between a single set of codeblocks (```). While you can write any text to think and plan before writing your JSON response, your JSON response must be the last thing you write, and it must be between a single set of codeblocks (```), and contain valid JSON, or it will not be parsed (which will be a critical error).

Your JSON response must be a dictionary that contains the following keys:

## `normalization_decisions` (list)
A list documenting key normalization decisions:
- `decision` (str): Description of the normalization
- `original_terms` (list[str]): The original terms that were normalized
- `normalized_to` (str): The final normalized term
- `affected_node_ids` (list[str]): Node IDs affected

## `nodes` (list)
A list of ontology node objects, where each node has:
- `id` (str): A unique identifier (e.g., "ont_1", "ont_2", etc.)
- `label` (str): A short, normalized label for the concept
- `description` (str): A fuller description of what this concept represents
- `category` (str): A category for the node (e.g., "method", "metric", "phenomenon")
- `abstraction_level` (str): One of "specific", "intermediate", or "abstract"
- `parents` (list[object]): List of parent relationships. Each object has two fields: `parent_id` (the parent node id) and `relation` (one of "is_a", "type_of", or "part_of"). For example: `[{"parent_id": "ont_1", "relation": "is_a"}]`. Omit this field entirely for root nodes (no parents). A node can have multiple parents to represent belonging to multiple categories.
- `aliases` (list[str]): Alternative names found in the observations. Omit this field if there are no aliases.


# Example Output
Here is a simplified example. Notice how the hierarchy has 3-4 levels of depth, with intermediate nodes bridging leaf concepts to root concepts. The intermediate nodes (ont_2, ont_3, ont_7, ont_8) were created using domain expertise — they do not need to appear in the input triplets.

```
{
    "normalization_decisions": [
        {
            "decision": "Merged various terms for large language models into a single concept",
            "original_terms": ["LLM", "large language model", "foundation model", "GPT-like model"],
            "normalized_to": "Large Language Model",
            "affected_node_ids": ["ont_3"]
        }
    ],
    "nodes": [
        {
            "id": "ont_1",
            "label": "AI Model",
            "description": "Artificial intelligence models and systems",
            "category": "system",
            "abstraction_level": "abstract",
            "aliases": ["AI system", "artificial intelligence model"]
        },
        {
            "id": "ont_2",
            "label": "Language Model",
            "description": "Models trained on text data to understand and generate natural language",
            "category": "system",
            "abstraction_level": "intermediate",
            "parents": [{"parent_id": "ont_1", "relation": "is_a"}]
        },
        {
            "id": "ont_3",
            "label": "Large Language Model",
            "description": "Language models with billions of parameters trained on large text corpora",
            "category": "system",
            "abstraction_level": "intermediate",
            "parents": [{"parent_id": "ont_2", "relation": "is_a"}],
            "aliases": ["LLM", "large language model", "foundation model"]
        },
        {
            "id": "ont_4",
            "label": "GPT-4",
            "description": "OpenAI's GPT-4 large language model",
            "category": "system",
            "abstraction_level": "specific",
            "parents": [{"parent_id": "ont_3", "relation": "is_a"}]
        },
        {
            "id": "ont_5",
            "label": "Claude-3.5-Sonnet",
            "description": "Anthropic's Claude 3.5 Sonnet model",
            "category": "system",
            "abstraction_level": "specific",
            "parents": [{"parent_id": "ont_3", "relation": "is_a"}]
        },
        {
            "id": "ont_6",
            "label": "Prompting Technique",
            "description": "Methods for structuring inputs to language models to elicit desired behaviors",
            "category": "method",
            "abstraction_level": "abstract",
            "aliases": ["prompting method", "prompt engineering"]
        },
        {
            "id": "ont_7",
            "label": "Reasoning Prompting",
            "description": "Prompting techniques that elicit step-by-step or structured reasoning",
            "category": "method",
            "abstraction_level": "intermediate",
            "parents": [{"parent_id": "ont_6", "relation": "is_a"}]
        },
        {
            "id": "ont_8",
            "label": "Chain-of-Thought Prompting",
            "description": "A prompting technique that elicits step-by-step reasoning from language models",
            "category": "method",
            "abstraction_level": "specific",
            "parents": [{"parent_id": "ont_7", "relation": "is_a"}],
            "aliases": ["CoT prompting", "CoT"]
        }
    ]
}
```


# Important Notes
- You must return a JSON response, and it must be valid JSON, or it will not be parsed.
- **Focus on building a useful hierarchy** - the main goal is to organize concepts at different abstraction levels so they can be normalized before building a causal graph.
- **Only use hierarchical relations** - stick to "is_a", "type_of", and "part_of". Do NOT include causal relations like "increases", "decreases", "enables", etc.
- **PRESERVE SPECIFIC CONCEPTS AS LEAF NODES** - This is critical! Keep specific model names (GPT-4, Claude, LLaMA), method names (LoRA, DPO, RLHF), dataset names (SQuAD, MMLU), and metric names (BLEU, F1) as individual leaf nodes. Only merge when the exact same concept appears with different syntax/formatting.
- **Normalize leaf nodes conservatively** - Only merge concepts that are syntactic variations of the exact same thing (e.g., "GPT-4" and "gpt4"). Do NOT merge related but distinct concepts (e.g., "GPT-3" and "GPT-4" must remain separate).
- **Normalize aggressively but meaningfully the further up the hierarchy** - merge concepts that are semantically equivalent, but preserve important distinctions.
- **100% concept coverage** - Every `source_concept` and `target_concept` string from the input triplets must map to an ontology node. If you normalized or merged a concept, document it in `normalization_decisions`.
- **The hierarchy should be balanced** - avoid very deep narrow branches or very flat wide structures.
- **Create intermediate nodes using domain expertise** - The input triplets mostly contain leaf-level concepts. You must invent intermediate and abstract grouping nodes to build a proper multi-level taxonomy. A leaf node connected directly to a root node is a sign of a missing intermediate layer.
- **Apply valence-aware grouping when appropriate** - For `outcome` / `metric` / `phenomenon` concepts whose siblings under a shared parent have a clear mix of good / bad / neutral natures, insert intermediate valence-grouping nodes ("Beneficial X" / "Adverse X" / "Neutral X") between the parent and the leaves. See "Valence-aware grouping for outcome-like concepts" in Section 2.
\end{lstlisting}

\subsection{Relation Normalization}
\label{app:prompt-relnorm}

Collapses the corpus's free-text relation labels onto a small canonical set (\S\ref{sec:impl-relnorm}).

\lstinputlisting[style=prompt]{prompts/relation_normalization.j2}

\subsection{Hypothesis and Theory Generation}
\label{app:prompt-theorizer}

We do not reproduce these two prompts. Both are adapted from Jansen et al.~\cite{284981954}, and the originals are published with their system. The changes are described in \S\ref{sec:impl-hypothesis} and \S\ref{sec:impl-theory}. The generator receives the researcher's curated canvas, or a weighted sample of the full graph, in place of a fixed retrieved corpus, and the theory prompt is given the researcher's selected hypotheses as fixed text to build a theory around instead of generating its own.

\subsection{Paper Question Answering}
\label{app:prompt-askpaper}

Backs the agent's \texttt{ask\_paper} tool (Table~\ref{tab:agent-tools}). The paper's full text is retrieved first and rendered into the prompt.

\begin{lstlisting}[style=prompt]
Answer a question using the provided scientific paper.
Your response should be a JSON object with the following fields:
  - answer: The answer to the question. The answer should use concise language, but be comprehensive. Only provide answers that are objectively supported by the text in paper.
  - excerpts: A list of string of one or more *EXACT* text spans extracted from the paper that support the answer. Return between at most five spans, and no more that 500 words. Make sure to cover all aspects of the answer above.

<Start of paper>

Title: {{ title }}

{{ full_text }}

<End of paper>

Given the information above, please answer the question: "{{ question }}"
\end{lstlisting}

\subsection{Chat Agent System Prompt}
\label{app:prompt-agent}

The system prompt held constant across every turn of a chat session (\S\ref{sec:impl-agent}). The canvas snapshot described in that section is not part of it; it is prepended to each user message instead, so that this text stays byte-identical across turns and can be cached between requests.

\lstinputlisting[style=prompt, breakatwhitespace=true]{prompts/chat_agent_system_prompt.unwrapped.txt}

\end{document}